\documentclass[journal=jctcce,manuscript=article]{achemso}
\setkeys{acs}{articletitle = true}

\mciteErrorOnUnknownfalse
\usepackage[version=3]{mhchem}
\usepackage{subfigure}
\usepackage[T1]{fontenc}
\usepackage{graphicx}
\usepackage[table,equation]{xcolor}
\usepackage{caption}
\makeatletter
\usepackage{bm}
\makeatother

\title{Semi-Automated Creation of Density Functional Tight Binding Models Through Leveraging Chebyshev Polynomial-based Force Fields}

\author{Nir Goldman} 
  \affiliation{Physical and Life Sciences Directorate, Lawrence Livermore National Laboratory, Livermore, California 94550, United States}
  \alsoaffiliation{Department of Chemical Engineering, University of California, Davis, California 95616, United States}
    \email{ngoldman@llnl.gov}
   \phone{+1-925-422-3994}
\author{Kyoung E. Kweon}
  \affiliation{Physical and Life Sciences Directorate, Lawrence Livermore National Laboratory, Livermore, California 94550, United States}
 \author{Babak Sadigh}
   \affiliation{Physical and Life Sciences Directorate, Lawrence Livermore National Laboratory, Livermore, California 94550, United States}
 \author{Tae Wook Heo}
   \affiliation{Physical and Life Sciences Directorate, Lawrence Livermore National Laboratory, Livermore, California 94550, United States}
\author{Rebecca K. Lindsey}
  \affiliation{Physical and Life Sciences Directorate, Lawrence Livermore National Laboratory, Livermore, California 94550, United States}
\author{C. Huy Pham}
  \affiliation{Physical and Life Sciences Directorate, Lawrence Livermore National Laboratory, Livermore, California 94550, United States}
\author{Laurence E. Fried}
  \affiliation{Physical and Life Sciences Directorate, Lawrence Livermore National Laboratory, Livermore, California 94550, United States}
\author{B\'alint Aradi}
  \affiliation{Bremen Center for Computational Materials Science, Universit\"at Bremen, P.O.B. 330440, D-28334, Bremen, Germany}
\author{Kiel Holliday}
  \affiliation{Physical and Life Sciences Directorate, Lawrence Livermore National Laboratory, Livermore, California 94550, United States}
\author{Jason R. Jeffries}
  \affiliation{Physical and Life Sciences Directorate, Lawrence Livermore National Laboratory, Livermore, California 94550, United States}
\author{Brandon C. Wood}
  \affiliation{Physical and Life Sciences Directorate, Lawrence Livermore National Laboratory, Livermore, California 94550, United States}
  
\date{\today}
\begin{document}

\newpage
\begin{abstract}

\end{abstract}

Density Functional Tight Binding (DFTB) is an attractive method for accelerated quantum simulations of condensed matter due to its enhanced computational efficiency over standard Density Functional Theory approaches. However, DFTB models can be challenging to determine for individual systems of interest, especially for metallic and interfacial systems where different bonding arrangements can lead to significant changes in electronic states. In this regard, we have created a rapid-screening approach for determining systematically improvable DFTB interaction potentials that can yield transferable models for a variety of conditions. Our method leverages a recent reactive molecular dynamics force field where many-body interactions are represented by linear combinations of Chebyshev polynomials. This allows for the efficient creation of multi-center representations with relative ease, requiring only a small investment in initial DFT calculations. We have focused our workflow on TiH$_2$ as a model system and show that a relatively small training set based on unit-cell sized calculations yields a model accurate for both bulk and surface properties. Our approach is easy to implement and can yield accurate DFTB models over a broad range of thermodynamic conditions, where physical and chemical properties can be difficult to interrogate directly and there is historically a significant reliance on theoretical approaches for interpretation and validation of experimental results.

\newpage

\section{Introduction}

Atomistic computer simulations are frequently used as an independent route to determining physical and chemical properties for complex systems, including material synthesis\cite{Irle13}, highly strained conditions\cite{Steele_VRDAC_2020}, and surface chemistry \cite{Mavrikakis_CO-Pd_2021}. These calculations can provide predictions of physical properties and reaction mechanisms, which can make experiments more tractable by aiding in their interpretation and helping to narrow the number of different conditions to investigate. Accurate modeling of the breaking and forming of chemical bonds in condensed phases frequently requires the use of quantum theories such as Kohn-Sham Density Functional Theory (DFT). DFT remains one of the most widely used theoretical methods in condensed matter physics, computational chemistry, and materials science for prediction of material properties and chemical reactivity, including phase boundaries and thermal decomposition (e.g., Ref.~\citenum{Schwegler08}). DFT calculations, though, require immense computational effort per simulation time step and consequently are usually limited to picosecond time scales and nanometer system sizes. In contrast, many processes of interest have properties that can span orders of magnitude larger scales, including large-scale carbon heterocycle synthesis\cite{Kroonblawd19-NPAH}, the rational design of 3D materials\cite{Kulkarni_MOF_2020}, and defect formation and grain boundary interactions in crystalline systems\cite{Sliwa_Ta_2018}. Hence, there is a great need to explore methods that can harness the accuracy of DFT while yielding substantial improvements in computational speeds.

Density Functional Tight Binding (DFTB) is a semi-empirical quantum theory based on an expansion of the Kohn-Sham DFT total energy that holds promise as an alternate approach for gas-phase studies\cite{Tkatchenko_DFTB-NN} as well as condensed phase systems\cite{Dolgonos10,DFTB_h2o_2020}. DFTB implements a balance of approximate quantum mechanics and empirical functions that can allow for several orders of magnitude improvement in computational efficiency for a variety of materials and mixtures\cite{DFTB+}. Difficulty can arise in creating new DFTB interaction potentials or improving upon existing models due to the large number of hyper-parameters within the approach. In particular, the approximate Hamiltonian in DFTB can be sensitive to the confining potential chosen for the wavefunction and electron density\cite{Aradi17} (Section~\ref{subsec:dftb}), and in general there does not exist a predefined recipe for how to choose these parameters nor how to explore that specific phase space. The Hamiltonian parameters in turn are closely coupled to the empirical repulsive energy, which itself has a wide variety of options in terms of functional form and data to be fit\cite{Goldman15-fm}. In addition, the repulsive energy is usually taken to be strictly pairwise (two-center), though a number of systems can require many-body terms as well for accurate predictions\cite{Goldman12b}. \textcolor{black}{The generation of training data can be problematic for high-Z or magnetic materials,\cite{Soderlind15} where sufficient DFT-MD data can be too computationally demanding to compute and being able to leverage small MD training sets would be advantageous. In addition, machine learning approaches tend to be highly data intensive\cite{Cheng_water_2020}, which can pose difficulties for any method that leverages these techniques.} Thus, DFTB method development would be holistically improved through a more automatic method for parameterization, where candidate models could be screened rapidly and efficiently, thereby allowing the user to quickly determine an optimal model for their specific needs.

To this end, we propose a method for streamlining DFTB parameterization by leveraging the recently developed Chebyshev Interaction Model for Efficient Simulation (ChIMES)\cite{Lindsey17, Lindsey_AL_2020}. ChIMES is a many-body reactive force field model for molecular dynamics (MD) simulation based on linear combinations of Chebyshev polynomials. ChIMES MD models have been created for a number of systems under both unreactive and reactive conditions, including molten liquid carbon\cite{Lindsey17}, water under ambient and high pressure-temperature conditions\cite{Koziol17,Lindsey_H2O_2019}, high pressure C/O systems\cite{LCO-2020, Lindsey_CO_2020}, and detonating energetic materials\cite{Pham_HN3_2021}. The main advantage of the ChIMES formalism is that optimal parameters are determined extremely rapidly (e.g., within minutes for this effort) through solving a set of simultaneous linear equations. This in contrast to bond order methods (e.g., Refs.~\citenum{Tersoff88, reaxff-h2o-xchem}) and neural network force fields (e.g., Refs.~\citenum{Behler-perspective, Tkatchenko_DFTB-NN}) that generally require non-linear optimization (e.g., Levenberg-Marquardt), making determination of optimal parameter sets significantly more labor intensive and uncertain due to possible local minima in the search space. 

\begin{figure}[!htp]
\begin{center}
\includegraphics[scale=0.8]{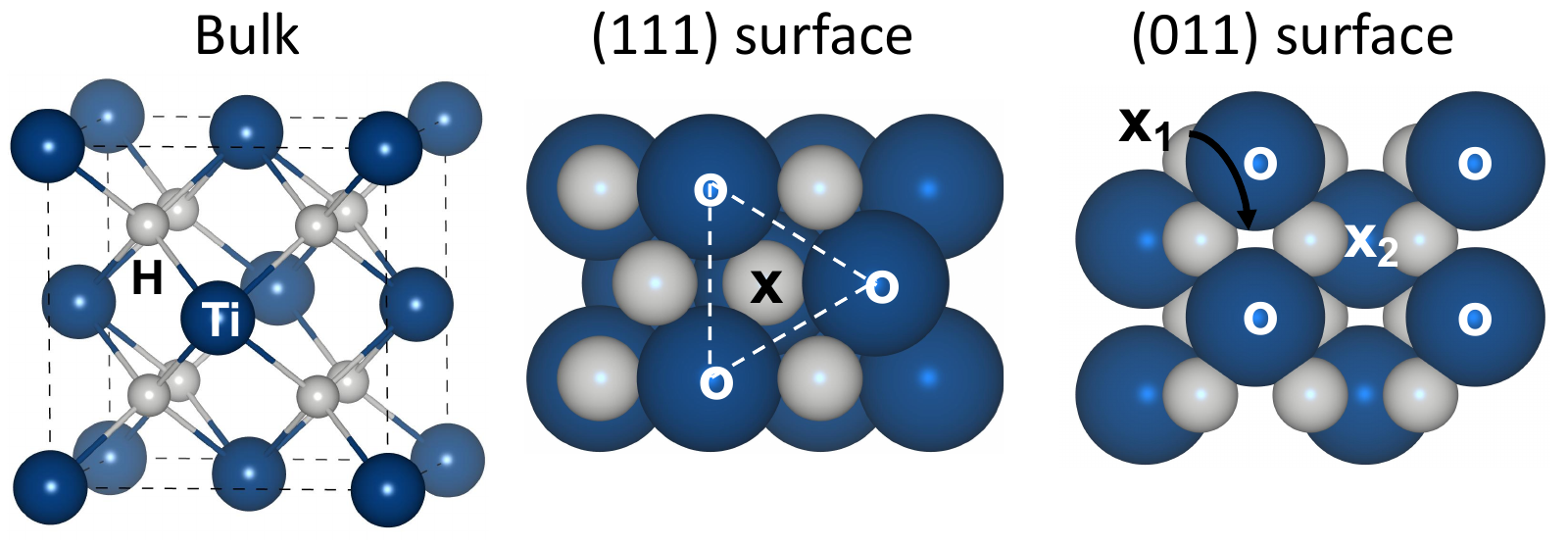}
\caption{\label{fig:tih2}  Pictures of TiH$_2$ bulk and surfaces. The left panel shows the bulk fcc lattice. The middle panel shows the (111) crystalline surface the Top (marked with an `O') and Hollow (`X') adsorption sites indicated. The right panel shows the (011) crystalline surface with the Top (`O'), Bridge-1 (`$\mathrm{X}_1$') and Bridge-2 (`$\mathrm{X}_2$') sites indicated.}
\end{center}
\end{figure}

In this work, we use the metal TiH$_2$ for our model determination. TiH$_2$ has a number of possible applications areas, including in hydrogen storage alloys\cite{Kitabayashi_MgH2-TiH2}, and as a superconducting material\cite{Shanavas_TiH2_2016}, a thermal energy storage material\cite{Shepard2016}, an intermediary to produce fine Ti metal powders\cite{Kovalev2010,Sandim2005}, and as a blending agent to create porous aluminum foams\cite{Peng_TiH2_blends_2016}. It exhibits face centered cubic (fcc) symmetry in its ground state, with the (111) surface generally exhibiting the lowest surface energy (Figure~\ref{fig:tih2}). Small scale TiH$_2$ calculations are relatively tractable with DFT, facilitating thorough validation of any developed DFTB parameter set. \textcolor{black}{However, DFT calculations are too computationally cumbersome to model the grain boundaries and density of crystalline defects that likely play a significant role in material performance in these application areas. These studies would thus greatly benefit from a high throughput quantum approach that can determine larger-scale effects that are challenging to determine from DFT alone.}  \textcolor{black}{In addition, study of TiH$_2$ serves as a starting point for determining (de)hydriding reactions in similar metal hydrides such as ZrH$_2$, which have a number of uses such as in solid propellants\cite{Yang_ZrH2_2020} and nuclear fuel cladding tubes\cite{Heo_ZrH2_2019}. The modeling framework developed here is general and can easily be applied to magnetic alloys such as Fe/Ni mixtures, which have use in magnetic refrigeration systems, magnetic recording devices, and other applications.\cite{Fe-Ni_alloy_2016}}

We begin with a brief discussion of our DFT calculations and the DFTB method, followed by details of the ChIMES formalism. We then experiment with different options for the DFTB quantum potential energy and charge transfer terms, and indicate optimal choices for our system. Next, we proceed through our workflow for DFTB model selection, beginning with the search for optimal values for the DFTB confining potentials, followed by the same for the ChIMES polynomials orders, and finally some discussion of optimal least-squares fitting practices. In all cases, we deliberately use a relatively small DFT training set as a somewhat stringent test for our results, and to assess the possibility of fitting to sparse data.

\section{Methods}

\subsection{DFT Calculations}
All DFT calculations were performed with the Vienna ab initio Simulation Package (VASP)\cite{vasp,vasp2,vasp3} using projector-augmented wave function (PAW) pseudopotentials\cite{Bloechl94,Kresse99} and the Perdew-Burke-Ernzerhof exchange correlation functional (PBE)\cite{Perdew:1996}. We found our results to be converged with a planewave cutoff of 400~eV and an energy convergence criteria of $10^{-6}$~eV, both of which were used for the results reported here. Fourth order Methfessel-Paxton smearing\cite{Methfessel-Paxton} was used with a value of 0.13~eV for all geometry and cell lattice optimizations in order to ensure energy convergence without dependence on the electronic smearing temperature. The Mermin functional\cite{Mermin65} with the same electronic temperature was used for all MD calculations in order to avoid spurious forces due to possible negative occupation numbers from the Methfessel-Paxton approach. Brillouin Zone sampling for all TiH$_2$ unit cell calculations was performed with a $10\times10\times10$ k-point mesh, whereas we used a mesh of $5\times5\times5$ for 32 formula unit (96 atom) bulk calculations. We used system sizes of 168 atoms/7~layers for the (001) surface, 144 atoms/6~layers for the (011) surface, and 192 atoms/8~layers for the (111) surface, each with a vacuum of 20$\,$\AA$\,$ and a k-point mesh of $5\times5\times1$ in the direction of the surface. 

Our DFT training set consisted of molecular dynamics simulations of unit cell configurations (12 atoms total), run for 5 ps at 400 K with simulation cells initially optimized to pressures of $-8$, $0$, $10$, and $100$~GPa. All MD calculations were run in the constant temperature and volume ($NVT$) ensemble with Nos\'e-Hoover thermostat chains\cite{nose-1984,hoover-1985,nose_hoover_chains} and a timestep of 0.2~fs. The slightly elevated temperature and wide pressure range including negative pressure were chosen in order to yield a broad sampling of the underlying potential energy surface. Atomic forces and the diagonal of the stress tensor were then sampled from MD configurations at fixed time intervals of $\sim160$ fs in order ensure configurations were as statistically uncorrelated as possible. This yielded up to 30 MD snapshots for each pressure. In addition, in order to sample hyper- and hypo-coordinated configurations in the system, we included MD data for a unit cell with a single hydrogen interstitial or single vacancy site, each run for 5 ps. The starting configuration for these defect simulations were taken from the ground-state stoichiometric unit cell, with a single H atom removed or added in an octahedral interstitial site. This yielded a total of 153 unit cell-sized configurations for our training set.

\subsection{DFTB theory and calculations}
\label{subsec:dftb}

The formalism for DFTB with self-consistent charges (SCC) has been discussed in detail elsewhere\cite{DFTB+_scc,Koskinen09,DFTB3,DFTB+,DFTB+_current}. Briefly, the method assumes neutral, spherically symmetric, atom-centered charge densities $n_0$ and that the true charge density of the system is within a small perturbation, i.e., $n(\mathrm{r}) = n_0(\mathrm{r}) + \delta n$. In its most typical form, a second-order Taylor expansion of the Kohn-Sham total energy in $\delta n$ is employed to yield the following energy expression:

\begin{equation}
\label{eqn:dftb}
\begin{split}
E[\delta n] \approx & \sum_a f_a \left<\psi_a \left| \hat{T} + \hat{V}\left[n_0\right] \right| \psi_a \right> \\
& +\frac{1}{2} \int \int^{'} \left(\frac{\delta^2E_{xc}[n_0]}{\delta n \delta n^{'}} + \frac{1}{\left| \mathbf{r} - \mathbf{r^{'}}\right|}\right)\delta n \delta n^{'} \\
& -\frac{1}{2} \int \frac{n^{'}_0 n_0}{\left| \mathbf{r} - \mathbf{r^{'}} \right|} + E_{xc}[n_0] - \int V_{xc}[n_0]n_0 + E_{II}, \\
\end{split}
\end{equation}

\noindent where we have employed a shorthand of $\int = \int dr$, $\int^{'} = \int dr^{'}$, $n_0\left(r\right) = n_0$, etc. The first line in Equation~\ref{eqn:dftb} is the band structure energy:

\begin{equation}
E_\mathrm{BS}[n_0] = \sum_a f_a \left< \psi_a \left| H[n_0] \right| \psi_a \right >, 
\label{eq:e_bs}
\end{equation}

\noindent where the Hamiltonian $H$ does not contain any charge transfer terms. $E_{\mathrm{BS}}$ is thus computed via sum over occupied electronic states determined from diagonalization of the approximate DFTB Hamiltonian. The Hamiltonian and overlap matrices used in $E_\mathrm{BS}$ are determined from pre-tabulated Slater-Koster (SK) tables derived from calculations with a minimal, non-orthogonal basis set. 

Prior to SK tabulation, both the electronic wave functions and electron density are subjected to separate confining or compression potentials. Most commonly, the confining potential has the general form of: 

\begin{equation}
V_{conf}(r) = \left( \frac{r}{R_\chi} \right)^\beta.
\label{eq:vconf}
\end{equation}

\noindent The exponent $\beta$ is usually set to a value of 2, and $R_\chi = R_\psi$ or $R_n$, where $R_\psi$ corresponds to wavefunction compression and $R_n$ to density compression. Our past experimentation with larger values of $\beta$  yielded negligible difference in the final outcome, though this could be investigated systematically in future efforts. The compression potentials force the wavefunction/electron density to zero at relatively large distances from the nuclei, which has been shown to improve transferability of the SK tabulations\cite{DFTB+_scc}. Choice of a short wavefunction compression improves transferability by removing the overly diffuse parts of the orbitals. However, this generally comes at the expense of the accuracy of the electron density at longer distances, which can affect the description of non-covalent interactions\cite{Elstner_SE_review}. Hence, the standard convention is to compute the wavefunction and density compression separately, with $R_n > R_\psi$. 

The potential energy in the DFTB Hamiltonian is generally expressed in one of two forms. In its original formulation\cite{Porezag95}, the Hamiltonian is taken to be a superposition of potentials centered on each atom, which for the purposes of this study we label as the ``\textcolor{black}{DFTB2-pot}'' model:

\begin{equation} 
\hat{H}^{{AB}}_{{\mu \nu}_{Tr1}} = \left<\phi^A_\mu \middle| \hat{T} + \hat{V}(n_A) + \hat{V}(n_B) \middle| \phi^B_\nu \right>
\label{eq:h_tr1}
\end{equation}

\noindent Here, $A$ and $B$ correspond to the atomic index, and $\mu$ and $\nu$ to an atomic orbital of specific angular momentum. This is akin to an N-body expansion of the quantum mechanical potential energy that is truncated after the one-center or one-body terms, only. 

A more common DFTB Hamiltonian involves a superposition of electronic densities instead, which we label as the ``\textcolor{black}{DFTB2-dens}'' model: 

\begin{equation} 
\hat{H}^{AB}_{{\mu \nu}_{Tr2}} = \left<\phi^A_\mu \middle| \hat{T} + \hat{V}(n_A  + n_B) \middle| \phi^B_\nu \right>
\label{eq:h_tr2}
\end{equation}

\noindent In this case, the electron density centered on atoms $A$ and $B$ is summed before evaluating the potential energy. This amounts to a two-body expansion, where the quantum mechanical potential energy includes both one and two-center terms. 

The second line from Equation~\ref{eqn:dftb} is the energy from second order charge fluctuations:

\begin{equation}
\label{eq:e2_coul}
E^2_\mathrm{Coul} =  \frac{1}{2} \int \int^{'} \left(\frac{\delta^2E_{xc}[n_0]}{\delta n \delta n^{'}} + \frac{1}{\left| \mathbf{r} - \mathbf{r^{'}}\right|}\right)\delta n \delta n^{'}.
\end{equation}

\noindent $E^2_\mathrm{Coul}$ is generally referred to as the Coulombic energy, and is evaluated self-consistently using the Hubbard $U$ parameter\cite{DFTB+_scc,Koskinen09}. Hubbard $U$ values for each atom type are precomputed with the uncompressed basis set as the difference between the ionization energy and electron affinity. 

In addition to these models, there exists a more recent DFTB energy expression, where third-order density fluctuation terms are included in the Kohn-Sham energy expansion\cite{Gaus12,DFTB3}.  This yields an additional charge-transfer term to Equation~\ref{eqn:dftb}, namely,

\begin{equation}
\label{eq:e3_coul}
E^3_\mathrm{Coul} = \frac{1}{6} \int \int^{'} \int^{''} \left(\frac{\delta^3 E_{xc}[n_0]}{\delta n\, \delta n^{'} \delta n^{''}}\right) \times \delta n\, \delta n^{'} \delta n^{''}.
\end{equation}

\noindent Evaluation of Equation~\ref{eq:e3_coul} requires the derivative of the Hubbard $U$ with respect to charge\cite{Gaus11}. This derivative can be determined numerically and for each angular momentum shell of each atom, separately. For our study, we include $E^3_\mathrm{Coul}$ only in conjunction with the density superposition potential (Equation~\ref{eq:h_tr2}), and label this interaction model as ``\textcolor{black}{DFTB3-dens}''. 

Finally, the last line in Equation~\ref{eqn:dftb} is called the repulsive energy:

\begin{equation}
\label{eqn:erep}
E_\mathrm{Rep} = -\frac{1}{2} \int \frac{n^{'}_0 n_0}{\left| \mathbf{r} - \mathbf{r^{'}} \right|} + E_{xc}[n_0] - \int V_{xc}[n_0]n_0 + E_{II}.
\end{equation}

\noindent The nomenclature derives from the ion-ion repulsion term, $E_{II}$, though it also contains Hartree double-counting and exchange-correlation terms. In practice, $E_{\mathrm{Rep}}$ is expressed as a short-ranged empirical function that covers first coordination shell or bonded interactions, only. Its parameters are usually fit to reproduce DFT or experimental data, and it can be either pair-wise\cite{Irle2020,Goldman15-fm} or contain multi-center interactions\cite{Goldman_DFTB_H_Pu,Dantana_OPV_2020}.

All DFTB calculations discussed within this work were performed with the DFTB+ code\cite{DFTB+,DFTB+_current}, using self-consistent charges (SCC)\cite{DFTB+_scc} and charge convergence criteria of $2.72 \times 10^{-5}$ eV ($10^{-6}$ au). Inclusion of an external van der Waals correction\cite{Rappe92,DFT-D} is beyond the scope of our present study. We have performed ``shell-resolved'' SCC calculations, where separate Hubbard $U$ parameters were determined for each orbital angular momentum shell. Hubbard $U$ derivatives were computed numerically for \textcolor{black}{DFTB3-dens} calculations, where we determined values of $-0.0328$, $-0.0840$, and $-0.05334$ for the $s$, $p$, and $d$-orbital interactions for Ti and a value of $-0.1857$ for the $s$-orbital for H. We have also applied an exponential charge damping factor for Ti-H interactions of 3.5, similar to previous efforts\cite{Gaus11}. The properties investigated here were largely insensitive to choice of either Hubbard $U$ derivative or damping factor values. The same electron thermal smearing and k-point mesh were used as described for our DFT calculations for each system. 

The ChIMES training set was determined by computing DFTB forces $(F)$ and diagonal stress tensor components $(\sigma)$ for each configuration with the chosen set of Hamiltonian parameters (i.e., $\{R_{\psi}\}$, $\{R_{n}\}$, \textcolor{black}{DFTB2-pot}, \textcolor{black}{DFTB2-dens}, or \textcolor{black}{DFTB3-dens} interaction) with zero values for those components from $E_\mathrm{Rep}$. These ``repulsive energy free'' results were then subtracted from the DFT values for those quantities, i.e., 
\begin{equation}
\begin{split}
F^{\tau*}_{\mathrm{Rep}_{\alpha_i}} = F^\tau_{\mathrm{DFT}_{\alpha_i}} - F^\tau_{\mathrm{QM,DFTB}_{\alpha_i}} \\ 
\sigma^{\tau*}_{\mathrm{Rep}_{\alpha\alpha}} = \sigma^\tau_{\mathrm{DFT}_{\alpha\alpha}} - \sigma^\tau_{\mathrm{QM,DFTB}_{\alpha\alpha}} 
\end{split}
\end{equation}

\noindent Here, $\tau$ corresponds to a specific MD configuration, $\alpha$ to the cartesian direction, and $i$ is the atomic index. The `*' is used to denote that the quantities being computed are part of the training set, and `QM,DFTB' refers to the quantum components of the DFTB calculation, i.e., only forces and stresses from $E_\mathrm{BS}$ and $E_\mathrm{Coul}$. Inclusion of configurational total energies generally resulted in minimal impact on quality of the $E_\mathrm{Rep}$ fit and were thus excluded from our training data, similar to previous efforts\cite{Goldman_DFTB_H_Pu,Dantana_OPV_2020}.  \textcolor{black}{This results in the following objective function:}
\color{black}{
\begin{equation}
    \label{eq:rmse}
    F_{obj} = \sqrt{\frac{1}{N_d}\left( \sum^M_{\tau=1} \sum^N_{i=1} \sum^3_{\alpha=1} \left[ F^{\tau}_{\mathrm{ChIMES}_{\alpha_i}} - F^{\tau*}_{\mathrm{Rep}_{\alpha_i}} \right]^2 + \sum^M_{\tau=1} \sum^3_{\alpha=1} \left[ \sigma^{\tau}_{\mathrm{ChIMES}_{\alpha\alpha}} - \sigma^{\tau*}_{\mathrm{Rep}_{\alpha\alpha}} \right]^2 \right)},
\end{equation}}

\noindent  \textcolor{black}{where $M$ is the total number of configurations in the training set, and $N_d$ is the total number of data entries (3$MN$ force components plus 3$M$ stress tensor components)}.
\color{black}
\subsection{ChIMES models and determination of parameters}

The design philosophy behind ChIMES involves mapping quantum mechanical energies onto linear combinations of many-body Chebyshev polynomials of the first kind. Chebyshev polynomials of the first kind have a number of desirable properties for creation of interatomic potential energy surfaces, including: (i) they are orthogonal (with respect to a weighting function) and can be generated recursively, allowing for basis set completeness and user defined complexity, (ii) higher order polynomials tend to have decreasing expansion coefficient values (due to their monic form), and (iii) they are ``nearly optimal'' (the error in an expansion will closely resemble a minimax polynomial). In addition, derivatives of Chebyshev polynomials of the first kind are are related to Chebyshev polynomials of the second kind, which themselves are orthogonal and can be generated recursively. This allows for easy and reliable determination of forces and stress tensor components for atomistic calculations. 

Briefly, the ChIMES total energy corresponds to an $n$-body expansion:

\begin{equation} 
E_{n_\mathrm{B}} = 
\sum^{n_\mathrm{a}}_{i_1}                      {}^1             \!E_{i_1} + 
\sum^{n_\mathrm{a}}_{i_1>i_2}                  {}^2             \!E_{i_1 i_2} + 
\sum^{n_\mathrm{a}}_{i_1>i_2>i_3}              {}^3             \!E_{i_1 i_2 i_3}  \\+ \dots + 
\sum^{n_\mathrm{a}}_{i_1>i_2\dots\ i_{n{_\mathrm{B}-1}}>i_{n{_\mathrm{B}}}}{}^{n_\mathrm{B}}\!E_{i_1 i_2 \dots i_n{_\mathrm{B}}},
\label{eq:genchimes}
\end{equation}

\noindent where $E_{n_\mathrm{B}}$ is the total ChIMES system energy, $n_\mathrm{B}$ is the maximum bodiedness, ${}^n\!E_{i_1 i_2 \dots\ i_n}$ is the $n$-body ChIMES energy for a given set of atoms with indices $\boldsymbol{i} = \{i_1, i_2,\ \dots\ , i_n\}$, and $n_\mathrm{a}$ is the total number of atoms in the system. The one-body energies, ${}^1\!E_{i_1}$, correspond to the atomic energy constants for each element type. 

The two-body (pairwise) energies are expressed as linear combinations of Chebyshev polynomials of the first kind:

\begin{equation} 
        {}^2\!E_{i_1 i_2} = f_{\mathrm{p}}\left(r_{i_1 i_2}\right) + f^{e_{i_1} e_{i_2}}_{\mathrm{c}}\left(r_{i_1 i_2}\right)
             \sum^{\mathcal{O}_{2}}_{m=1} C_m^{e_{i_1} e_{i_2}} T_m (s^{e_{i_1} e_{i_2}}_{i_1 i_2})
        \label{eqn:FF2B}
\end{equation}

\noindent In this case, $T_{m}\left(s^{e_{i_1} e_{i_2}}_{i_1 i_2}\right)$ represents a Chebyshev polynomial of order $m$, and $s^{e_{i_1} e_{i_2}}_{i_1 i_2}$ is the pair distance transformed to occur over the interval $[-1,1]$ using a Morse-like function\cite{WHBB-2009,WHBB-2011} (See Ref.~\citenum{Lindsey17} for details). Here, $s^{e_{i_1} e_{i_2}}_{i_1 i_2} \propto \mathrm{exp}\left(-r_{i_1 i_2}/\lambda_{e_1 e_2}\right)$ and $\lambda_{e_1 e_2}$ is an element-pair distance scaling constant, usually taken to be the peak position of the first coordination shell. $C_{m}^{e_{i_1} e_{i_2}}$ is the corresponding permutationally invariant coefficient for the interaction between atom types $e_{i_1}$ and $e_{i_2}$, taken from the set of all possible element types, $\{\boldsymbol{e}\}$. The term $f^{e_{i_1} e_{i_2}}_\mathrm{c}(r_{i_1 i_2})$ is a Tersoff cutoff function\cite{Tersoff88} which is set to zero beyond a maximum distance defined for a given $\{e_1, e_2\}$ pair set. In order to prevent sampling of $r_{i_1 i_2}$ distances below what is sampled in our DFT training set, we introduce use of a smooth penalty function $f_\mathrm{p} (r_{i_1 i_2})$. We refer the reader to previous work for additional details\cite{Lindsey_H2O_2019}.

We can now create a greater than two-body orthogonal basis set by taking products of the ${n \choose 2}$ unique constituent pairwise polynomials of the higher order terms. In other words, a three-body term has ${3 \choose 2} = 3$ pairs, which yields the following expression for the ChIMES three-body energy:

\begin{equation} 
{}^3\!E_{i_1 i_2 i_3} =       f^{e_{i_1} e_{i_2}}_{\mathrm{c}}\left(r_{i_1 i_2}\right)
                                f^{e_{i_1} e_{i_3}}_{\mathrm{c}}\left(r_{i_1 i_3}\right)
                                f^{e_{i_2} e_{i_3}}_{\mathrm{c}}\left(r_{i_2 i_3}\right)
                \sum^{\mathcal{O}_{3}}_{m=0}
                \sum^{\mathcal{O}_{3}}_{p=0}
                {\sum^{\mathcal{O}_{3}}_{q=0}}^\prime
                C^{e_{i_1} e_{i_2} e_{i_3}}_{mpq}
                                T_{m}\left(s^{e_{i_1} e_{i_2}}_{i_1 i_2}\right)
                                T_{p}\left(s^{e_{i_1} e_{i_3}}_{i_1 i_3}\right)
                                T_{q}\left(s^{e_{i_2} e_{i_3}}_{i_2 i_3}\right).
        \label{eqn:FF3B}
\end{equation}

\noindent We take a triple sum for the $i_1 i_2$, $i_1 i_3$, and $i_2 i_3$ polynomials over the hypercube up to $\mathcal{O}_{3}$, and include a single permutationally invariant coefficient for each set of powers and atom types, $C^{e_{i_1} e_{i_2} e_{i_3}}_{mpq}$. We use the primed sum to denote that only terms for which two or more of the $m,p,q$ polynomial powers are greater than zero are included in order to guarantee that three distinct atom-centers are evaluated.  The expression for ${}^3\!E_{i_1 i_2 i_3}$ also contains the $f_{\mathrm{c}}$ smoothly varying cutoff functions for each constituent pair distance. Penalty functions are not included in this case and instead are handled entirely by the two-body interaction.

Similarly, the four-body energy can be written as a product of the ${4 \choose 2} = 6$ unique pairwise interactions:

\begin{equation}
\begin{aligned}
{}^4\!E_{i_1 i_2 i_3 i_4} 
&= f^{e_{i_1} e_{i_2}}_c(r_{i_1 i_2}) f^{e_{i_1} e_{i_3}}_c(r_{i_1 i_3}) f^{e_{i_1} e_{i_4}}_c(r_{i_1 i_4}) f^{e_{i_2} e_{i_3}}_c(r_{i_2 i_3}) f^{e_{i_2} e_{i_4}}_c(r_{i_2 i_4}) f^{e_{i_3} e_{i_4}}_c(r_{i_3 i_4}) \\
& \times \sum_{m=0}^{{\cal O}_{\rm{4B}}} 
\sum_{p=0}^{{\cal O}_{\rm{4B}}}
\sum_{q=0}^{{\cal O}_{\rm{4B}}}
\sum_{r=0}^{{\cal O}_{\rm{4B}}}
\sum_{s=0}^{{\cal O}_{\rm{4B}}}
{\sum_{t=0}^{{\cal O}_{\rm{4B}}}}^{\prime} 
C^{e_{i_1} e_{i_2} e_{i_3} e_{i_4}}_{mpqrst} 
T_m (s^{e_{i_1} e_{i_2}}_{i_1 i_2})
T_p (s^{e_{i_1} e_{i_3}}_{i_1 i_3})
T_q (s^{e_{i_1} e_{i_4}}_{i_1 i_4})
T_r (s^{e_{i_2} e_{i_3}}_{i_2 i_3})
T_s (s^{e_{i_2} e_{i_4}}_{i_2 i_4})
T_t (s^{e_{i_3} e_{i_4}}_{i_3 i_4}).
\end{aligned}
\label{eq:ChIMES_4B}
\end{equation}

\noindent Again, the $C^{e_{i_1} e_{i_2} e_{i_3} e_{i_4}}_{mpqrst}$ correspond to permutationally invariant coefficients of linear combination. In this case, the primed sum indicates that three or more of the $m,~p,~q,~r,~s,~t$ polynomial orders are greater than zero. Thus far we have limited our studies to include up to four-bodied interactions\cite{Pham_HN3_2021}. However, the practice of generating higher-bodied terms through multiplication of the constituent pairwise polynomials can be generalized to include an $n \choose 2$-multiple sum over the $n$-body hypercube\cite{Lindsey_AL_2020}, which is the subject of future work. 

Optimal ChIMES parameters (the coefficients of linear combination) can then readily be determined through the overdetermined matrix equation $\boldsymbol{A} \boldsymbol{C} = \boldsymbol{B}_{\mathrm {rep}}$. The matrix $\boldsymbol{A}$ corresponds to the values of the requisite polynomials for a given training configuration, or in other words the derivatives with respect to the fitting coefficients. The column vectors $\boldsymbol{C}$ and $\boldsymbol{B}_{\mathrm {rep}}$ correspond to the linear ChIMES coefficients and the numerical values for the training set, respectively. This linear least-squares optimization problem can be solved for with any number of algorithms (Section~\ref{subsec:lsq}). 

\section{Results and Discussion}

\subsection{Tests of different forms of the DFTB interaction energy}

\textcolor{black}{In order to test each of our DFTB interaction models, we have initially set Ti compression radii to $R^{Ti}_\psi = 5.0$ au and $R^{Ti}_n = 14.0$ au. These are similar to established values for titanium, and such as the matsci-0-2\cite{matsci02}, trans3d-0-1\cite{trans3d}, and tiorg-0-1\cite{tiorg} parameter sets. For the entirety of our study, hydrogen compression radii were set to the same values as those for the miomod-hh-0-1 parameter set, namely, $R^{H}_\psi = 2.5$ au and $R^{H}_n = 3.0$ au.} We then generated the SK tables for all interactions. For this section, we have computed single-point properties based on the DFT optimized unit cell, e.g., without determining $E_\mathrm{Rep}$ and performing a geometry or lattice optimization. In doing so, we can probe the quality of each interaction model through direct one-to-one comparison to results for DFT computed eigenstates without having to be concerned with the quality of the repulsive energy parameterization.

\begin{figure}[!htp]
\begin{center}
\includegraphics[scale=1.0]{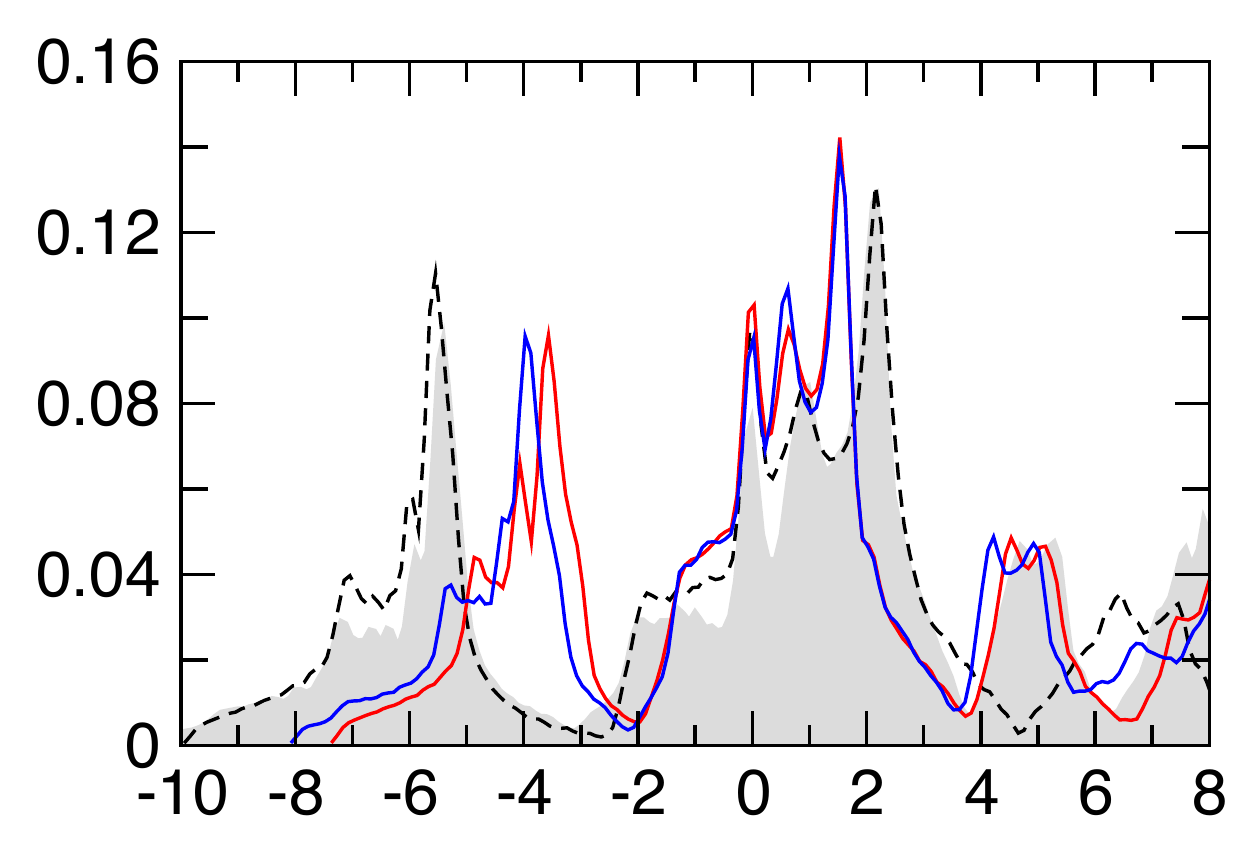}
\caption{\label{fig:edos} Total electronic density of states for bulk face centered cubic TiH$_2$. The dashed corresponds to the result from the \textcolor{black}{DFTB2-pot} Hamiltonian, the red line to \textcolor{black}{DFTB2-dens}, the blue line to \textcolor{black}{DFTB3-dens}, and the shaded area to DFT.  Each curve is centered around its Fermi energy. }
\end{center}
\end{figure}

Analysis of the total electronic density of states (DOS) reveals that all three interaction models yield similar curve shapes relative to DFT in terms of peak energy positions and their relative heights (Figure~\ref{fig:edos}). We have centered each DOS curve around its Fermi energy ($E_\mathrm{Fermi} = 0$) in order to facilitate comparison. The \textcolor{black}{DFTB2-pot} interaction yields the closest comparison to DFT in terms of the overall peak positions and heights, including eigenstates up to $\sim$3~eV. However, the agreement between states above this is significantly poorer, where \textcolor{black}{DFTB2-pot} completely omits the doublet at $~\sim$5~eV present in DFT.  In contrast, \textcolor{black}{DFTB2-dens} and \textcolor{black}{DFTB3-dens} show less accurate agreement with DFT for eigenstates below 3~eV, but improved agreement above this value where both models yield the 5~eV doublet.  

We have also determined the Mulliken atomic charges for Ti and H for the bulk system (Table~\ref{tab:charges}), with comparison to Bader charges determined from VASP. While determining a direct correspondence between Mulliken and Bader charges is not possible, we observe reasonable agreement between \textcolor{black}{DFTB2-dens} and \textcolor{black}{DFTB3-dens} with the results from DFT. In contrast, charges computed from \textcolor{black}{DFTB2-pot} are several orders of magnitude smaller and of the opposite sign to the other methods. We observe relatively small differences between \textcolor{black}{DFTB2-dens} and \textcolor{black}{DFTB3-dens}, either for the electronic DOS or for the atomic charges, with \textcolor{black}{DFTB2-dens} arguably yielding slightly improved agreement with the DOS from DFT. \textcolor{black}{The $E^3_\mathrm{Coul}$ interaction was developed in part to improve descriptions of polarizability in hydrogen-bonded systems\cite{Goyal_DFTB3_water_2014}, and consequently its effect could prove more substantial for highly polarizable or charged species.} Regardless, for the remainder of this work we will use the \textcolor{black}{DFTB2-dens} interaction model.

\subsection{Description of the $E_\mathrm{Rep}$ optimization workflow}

\begin{figure}[!htp]
\begin{center}
\includegraphics[scale=0.7]{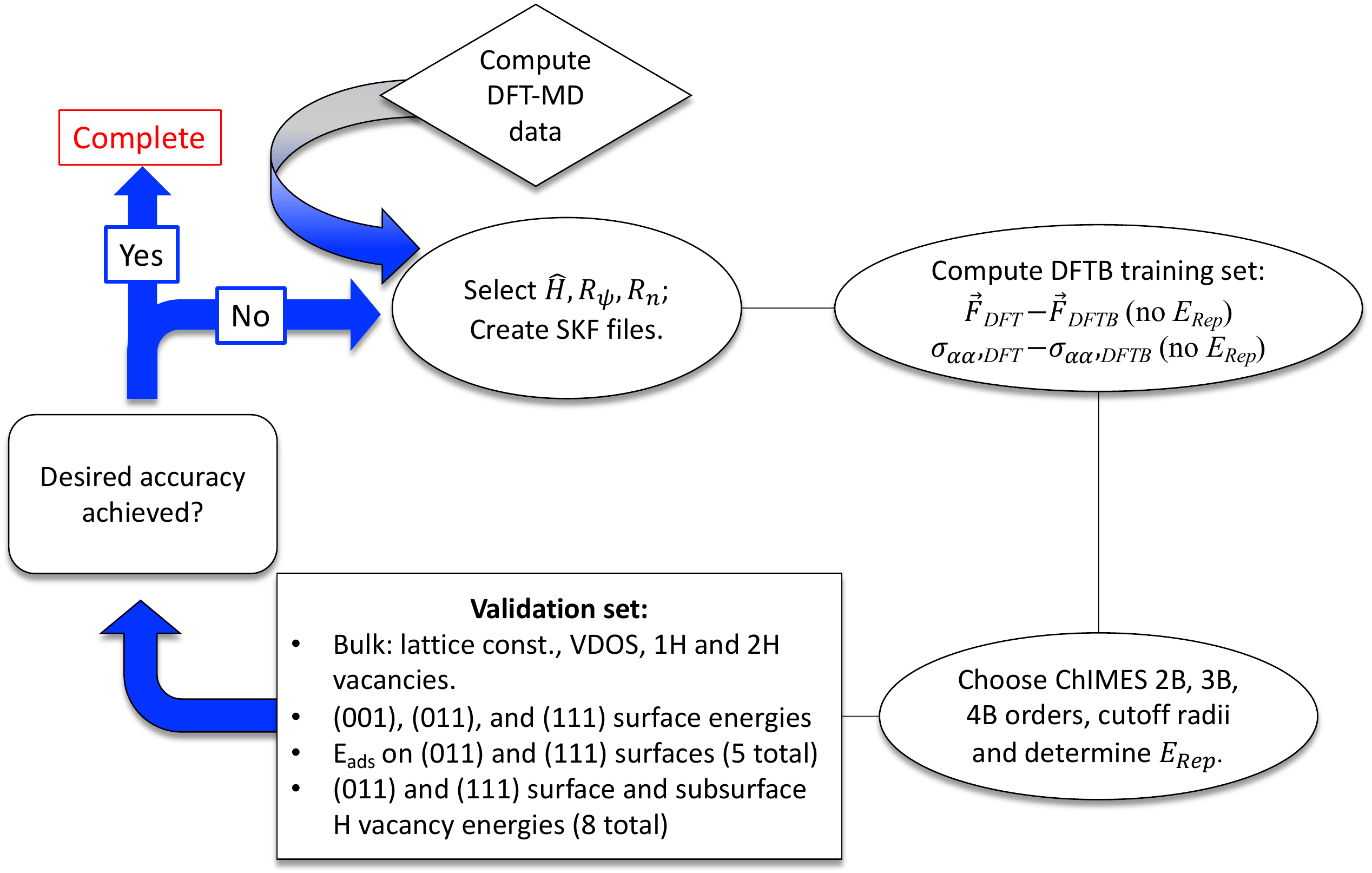}
\caption{\label{fig:flowchart} Flowchart for creation of DFTB $E_\mathrm{rep}$ models through ChIMES force field parameterization.}
\end{center}
\end{figure}

Given our choice of interaction potential of superposition of densities and second-order $E^2_\mathrm{Coul}$, we can now create a workflow for determination of optimal DFTB hyper-parameters (Figure~\ref{fig:flowchart}). This includes evaluation of $R_\psi$ and $R_n$ values, two-body or greater polynomial order for the ChIMES $E_\mathrm{Rep}$, and choice of linear optimization and regularization algorithms for calculation of the ChIMES coefficients. Our workflow is fast and efficient, with the most significant amount of time and CPU resources spent on computing the DFT training set. Each loop of our workflow has been automated, with the user being able to make decisions for how to proceed with hyper-parameter adjustments based on results for validation data.

Our initial DFT-MD simulations were run concurrently for 24 hours on four nodes each of a Linux cluster, with thirty-six 2.1 GHz Intel Xeon E5-2695 v4 cores and 128 gigabytes memory per node. This represented a usage of over 20,000 CPU hours. Generation of SK tables for chosen values of $R_\psi$ and $R_n$ run were on a single CPU and required approximately one hour of wall-clock time for each set. \textcolor{black}{Again, hydrogen compression radii were held at miomod-hh-0-1 values.} \textcolor{black}{For this effort, we choose to use a `global' compression radius for Ti, where all orbitals use the same confinement radius. This is consistent with previous Ti related parameterizations available for download\cite{matsci02,trans3d,tiorg}, all which use the same confinement radius for s, p, and d orbitals. There is some possibility that separate confinement for the vacant 4p orbital could yield improved accuracy\cite{Irle2020}, which is the subject of future work.} Calculation of the $E_\mathrm{Rep}$ training set was run using twelve OpenMP threads, requiring $\sim$30 minutes. Solving the linear least-squares problem for a single ChIMES parameterization required less than one minute on a single CPU in all cases, allowing for extremely rapid screening of different sets of parameters. All minimum and cutoff radii for the ChIMES $E_\mathrm{Rep}$ were set to include the first coordination shell sampled in our training set, only: $2.5 \leq r_{TiTi} \leq 3.5$ $\,$\AA$\,$ and $1.5 \leq r_{HTi}  \leq 2.5$ $\,$\AA$\,$. We use values of $\lambda_{TiTi} = 3.0\,$\AA$\,$ and $\lambda_{HTi} = 2.0\,$\AA$\,$ for the Morse-like coordinate transforms. H-H repulsive interaction were not sampled in our training set and were thus takes from the miomod-hh-0-1 parameter set. 

ChIMES parameterization was followed by calculation of validation data, which included results explicitly not present in our training data. The total validation set included the bulk lattice constant and hydrogen vacancy energy, surface energies of the (001), (011) and (111) crystal facets, hydrogen adsorption energies on the (011) and (111) surfaces, and (011) and (111) surface and sub-surface hydrogen vacancy energies. The (001) surface constructed for our calculations contained a contained a net dipole field due to an imbalance of H ions on the top vs. bottom surface for the stoichiometric system, and as such we did not include its hydrogen surface adsorption and vacancy energies in our validation. Calculation of the entire validation set was performed using twelve OpenMP threads and required approximately one hour to complete. Thus, each design loop in our workflow that included generation of SK tables and $E_\mathrm{Rep}$ training data took about two and half hours of wall clock time overall. 

\subsection{Sweep of $R_\psi$ and $R_n$ values}

For this effort, we have computed SK tables using confining radii values ranging from $3.2 \leq R^{Ti}_\psi \leq 5.0$ au and $6.0 \leq R^{Ti}_n \leq17.0$ au. Hydrogen compression radii were held at the previously mentioned values. We find that this range of radii spans those normally sampled in DFTB models for low-Z elements as well as transition metals. Smaller values of $R^{Ti}_\psi$ and $R^{Ti}_n$ resulted in some distortions to the electronic DOS without yielding additional accuracy for our validation tests. For this section of our study, we used a constant ChIMES 2-body polynomial order of 12 and a 3-body order of 8, similar to previous work\cite{Goldman_DFTB_H_Pu,Dantana_OPV_2020}. All optimization discussed in here were performed with the LASSO/LARS algorithm with regularization of $10^{-3}$ (see Section~\ref{subsec:lsq} for details). In doing so, we are able to rapidly create a reasonable $E_\mathrm{Rep}$  for our models, allowing us to assess the effects of different values of $R^{Ti}_\psi$ and $R^{Ti}_n$ independently.

We find that that there was small variation between results the root mean-square (RMS) errors for the training set, as well as for validation of the bulk lattice constant, vacancy energies, and (011) hydrogen surface adsorption energies. In addition, all models yielded the DFT-predicted surface energy ordering of $E_{111} < E_{011} < E_{001}$. Hence, we focus our validation discussion on the $E_{111}$ surface energy, the $\left(E_{001}/E_{111}\right)$ ratio, and the relative energies of adsorption on the (111) Top and Hollow surface sites, $\left(E^\mathrm{Top}_{111} -  E^\mathrm{Holl}_{111}\right)$.

\begin{figure}[!htp]
\begin{center}
\includegraphics[trim=1cm 2cm 1cm 2cm,scale=0.31]{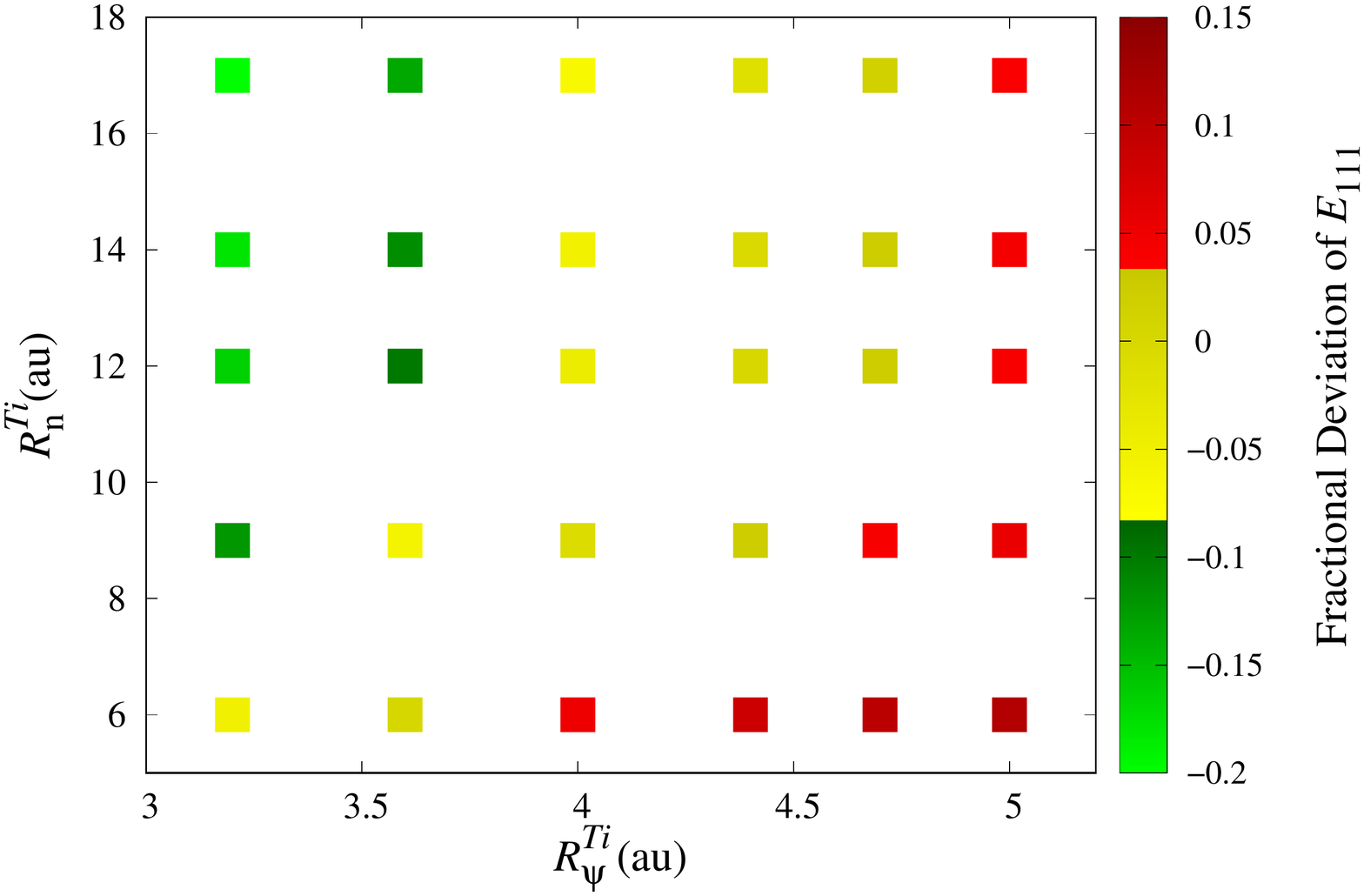}
\includegraphics[trim=1cm 2cm 1cm 2cm,scale=0.31]{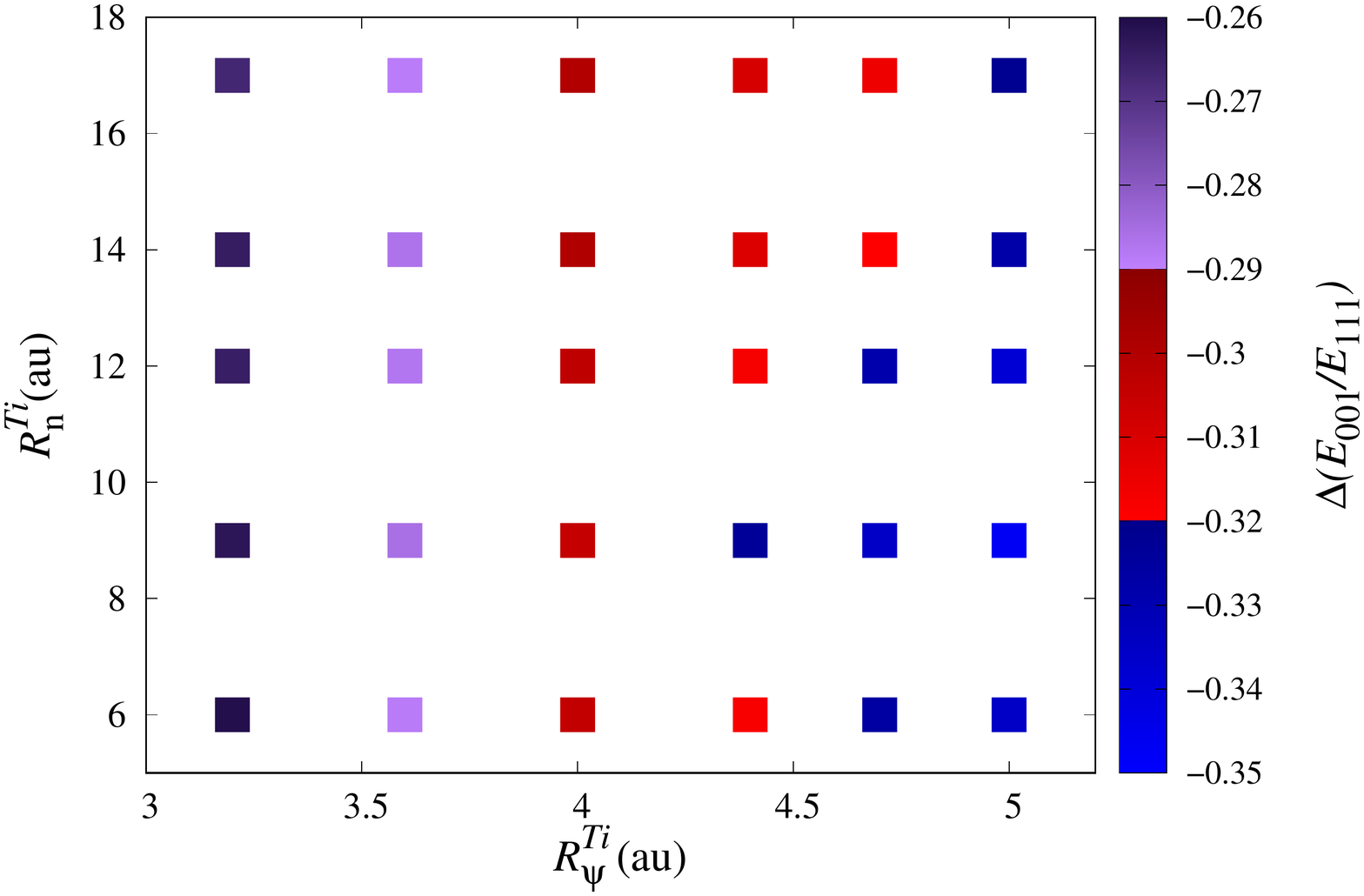}
\includegraphics[trim=1cm 2cm 1cm 2cm,scale=0.31]{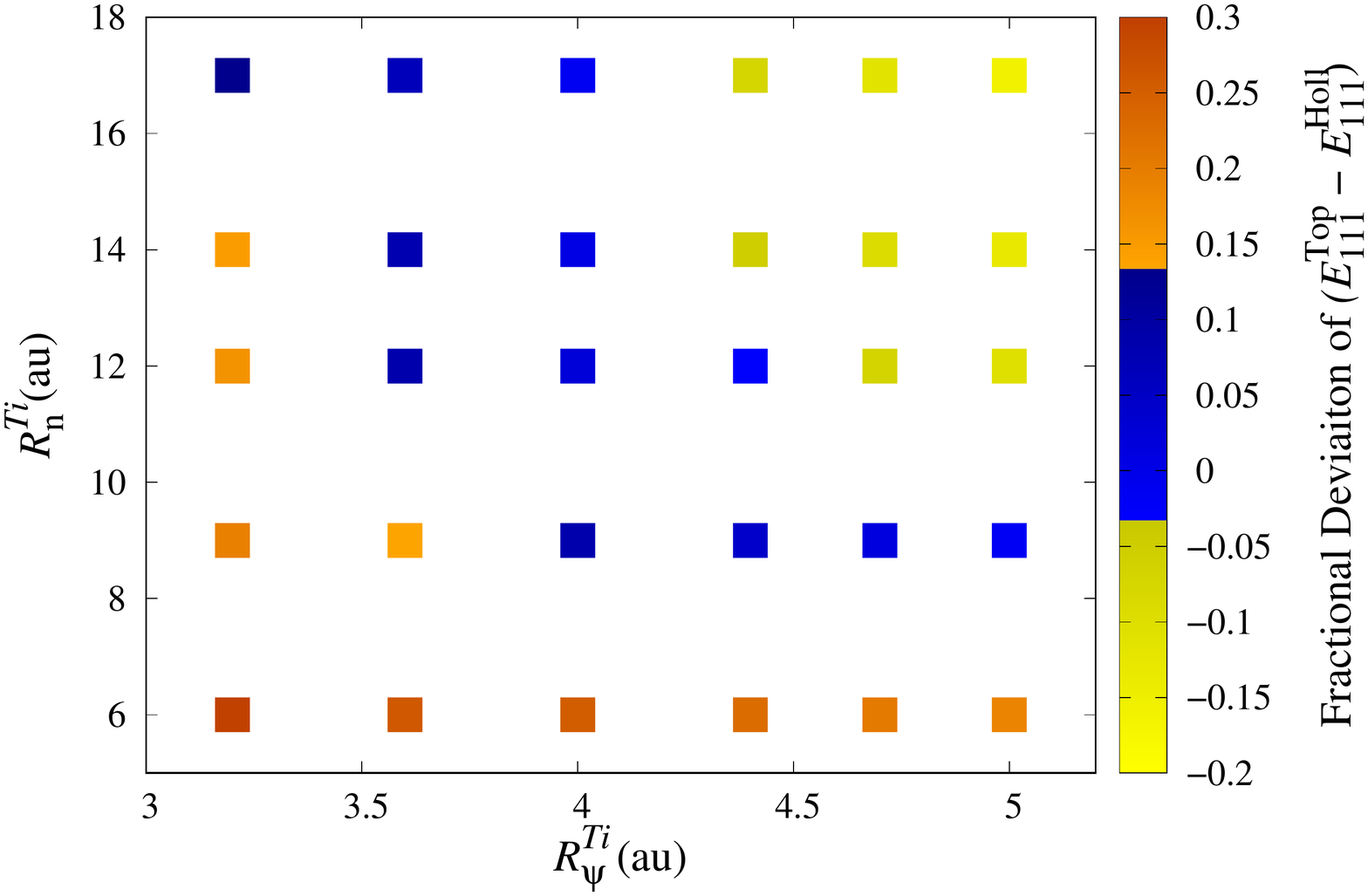}
\caption{\label{fig:radius_sweep} Results for sweep of values of $R^{Ti}_{\psi}$ and $R^{Ti}_{n}$. The top panel corresponds to the fractional deviation of the surface energy, $\left(E^{DFTB}_{111} - E^{DFT}_{111}\right)/E^{DFT}_{111}$, and the middle panel to the deviation of $\left(E_{001}/E_{111}\right)$ relative to DFT. The bottom panel corresponds to the fractional deviation of $\left(E^\mathrm{Top}_{111} -  E^\mathrm{Holl}_{111}\right)$, where we have shifted the results so that positive values indicate the correct ordering of $E^\mathrm{Top}_{111} >  E^\mathrm{Holl}_{111}$. A perfect matching with DFT in this case would yield a value of one. }
\end{center}
\end{figure}

Our analysis reveals an approximate linear correlation between \{$R^{Ti}_\psi$, $R^{Ti}_n$\} in terms of the accuracy of the $E_{111}$ energy (Figure~\ref{fig:radius_sweep}, top panel). The most accurate results for the $E_{111}$ surface energy appear to occur when the radii are approximately consistent with each other (either low or high). Small values of $R^{Ti}_\psi$ combined with long values of $R^{Ti}_n$ tend to over-predict $E_{111}$, where \{$R^{Ti}_\psi=3.2$, $R^{Ti}_n=17.0$\} predicts a surface energy that is $\sim$20\% too large. In contrast, larger values of $R^{Ti}_\psi$ combined with smaller values of $R^{Ti}_n$ tend to under-predict, where \{$R^{Ti}_\psi=5$, $R^{Ti}_n=6.0$\} yields a surface energy that is $\sim$11\% too low. 

In contrast, all of models created here under-predict $\left(E_{001}/E_{111}\right)$ relative to our DFT calculations (Figure~\ref{fig:radius_sweep}, middle panel). We observe ratios of 1.44 or so for small $R^{Ti}_\psi$ that monotonically decrease to values of $\sim$1.35 as $R^{Ti}_\psi$ is increased, whereas the DFT predicted value is 1.70. Our results indicate a much smaller dependence on choice of  $R^{Ti}_n$, where altering its value creates little change in the ratio for a given $R^{Ti}_\psi$.   We note that there can be strong dependence of the surface energies on choice of DFT functional (e.g., Ref.~\citenum{Thiel_Cu_surfaces}), although the relative energetic ordering tends to be consistent.

Finally, our models yield a variety of results for $\left(E^\mathrm{Top}_{111} -  E^\mathrm{Holl}_{111}\right)$ (Figure~\ref{fig:radius_sweep}, bottom panel), where DFT yields top site adsorption energies that are 0.68 eV higher than that for the hollow site. We found that predicting this energy difference was a particularly difficult challenge for our modeling effort. This is in part likely because Top site adsorption yields a hyper-coordinated Ti surface atom with H chemisorption, whereas Hollow site adsorption yields a more energetically favorable physisorption interaction with the H ion at larger distances from the surrounding surface Ti atoms. For this set of ChIMES parameters, we find that large $R^{Ti}_\psi$ and $R^{Ti}_n$  tend to reverse the energetic ordering of adsorption on these sites, with a top site energy that is 0.1 eV lower than that for the hollow site. In contrast, the results for $R^{Ti}_n = 6.0$ generally yielded the best agreement with DFT. These results show monotonic decreases with increasing $R^{Ti}_\psi$, with a value of 0.20 eV for $R^{Ti}_\psi = 3.2$ and a value of 0.13 eV for $R^{Ti}_\psi = 5.0$.

Based on these results, we choose to proceed with our SK tables generated from \{$R^{Ti}_\psi = 3.6$, $R^{Ti}_n = 6.0$\}. We find this combination yields a nearly identical $E_{111}$ to DFT with a value of 0.080 eV/\AA$^2$, and provides a relatively high $\left(E_{001}/E_{111}\right)$ ratio of 1.41 for this set of ChIMES parameters.  In addition, this model provides a reasonable prediction for the top and hollow site adsorption energies, with an energy difference of 0.17 eV. This model set was chosen over \{$R^{Ti}_\psi = 3.2$, $R^{Ti}_n = 6.0$\} due to the similarity in accuracy between these sets and in order to avoid any unforeseen distortions to the electronic states due to small confining radii.   

\subsection{Sweep of ChIMES polynomial orders}

Given our choice of \{$R^{Ti}_\psi = 3.6$, $R^{Ti}_n = 6.0$\}, we can now experiment with ChIMES polynomial orders in order to refine our model further. In this case, we have looped over our workflow by sampling the 2-body order a range of $8 \leq {{\cal O}_{\rm{2B}}} \leq 16$ and the 3-body order over a range of $0 \leq {{\cal O}_{\rm{3B}}} \leq 16$. Once again, we present results from LASSO/LARS optimization with regularization of $10^{-3}$. Each design loop in this case required only $\sim$1 hour (since SK table and $E_\mathrm{Rep}$ training data generation were complete), almost all of which was spent on validation calculations.

\begin{figure}[!htp]
\begin{center}
\includegraphics[trim=1cm 2cm 1cm 2cm,scale=0.31]{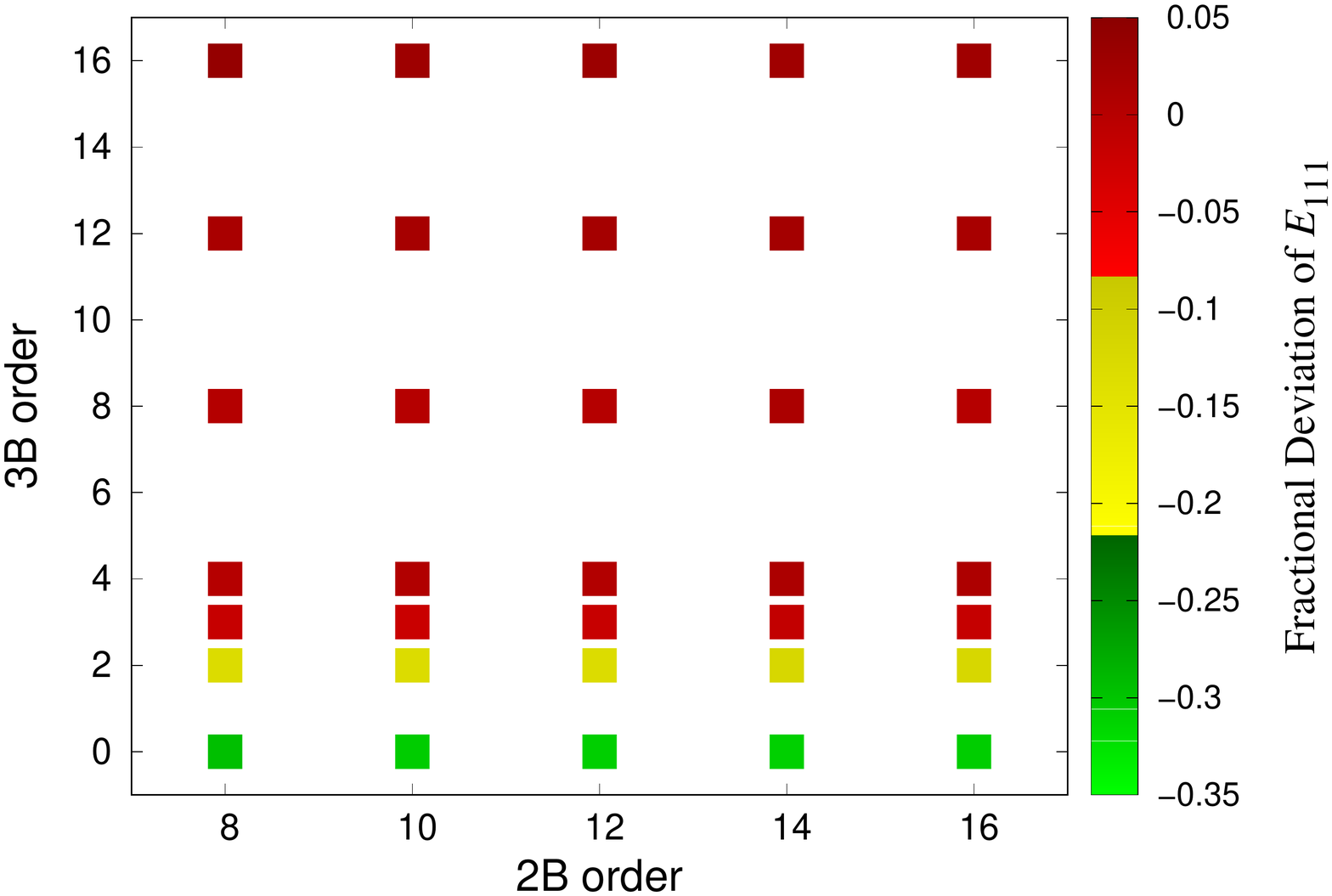}
\includegraphics[trim=1cm 2cm 1cm 2cm,scale=0.31]{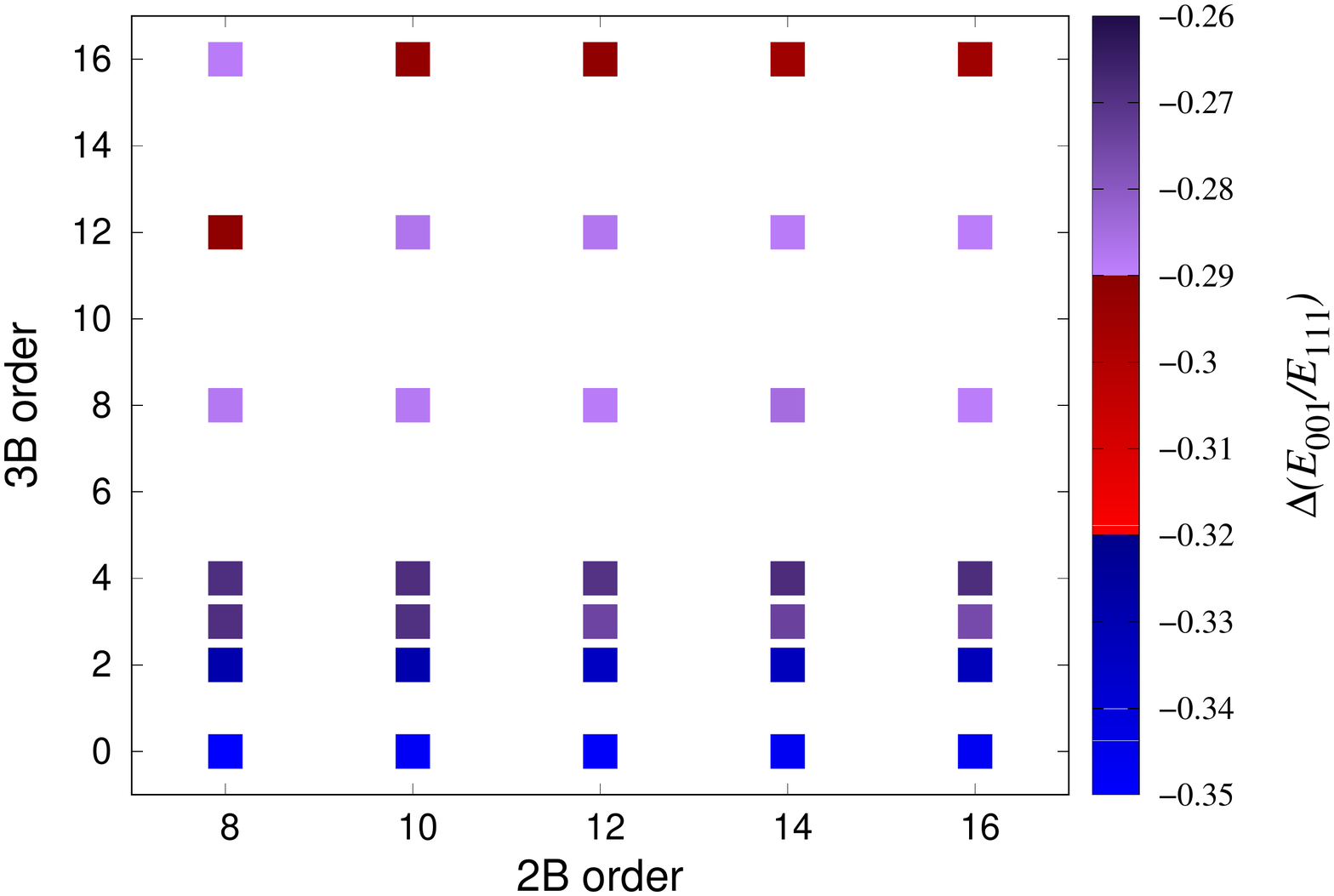}
\includegraphics[trim=1cm 2cm 1cm 2cm,scale=0.31]{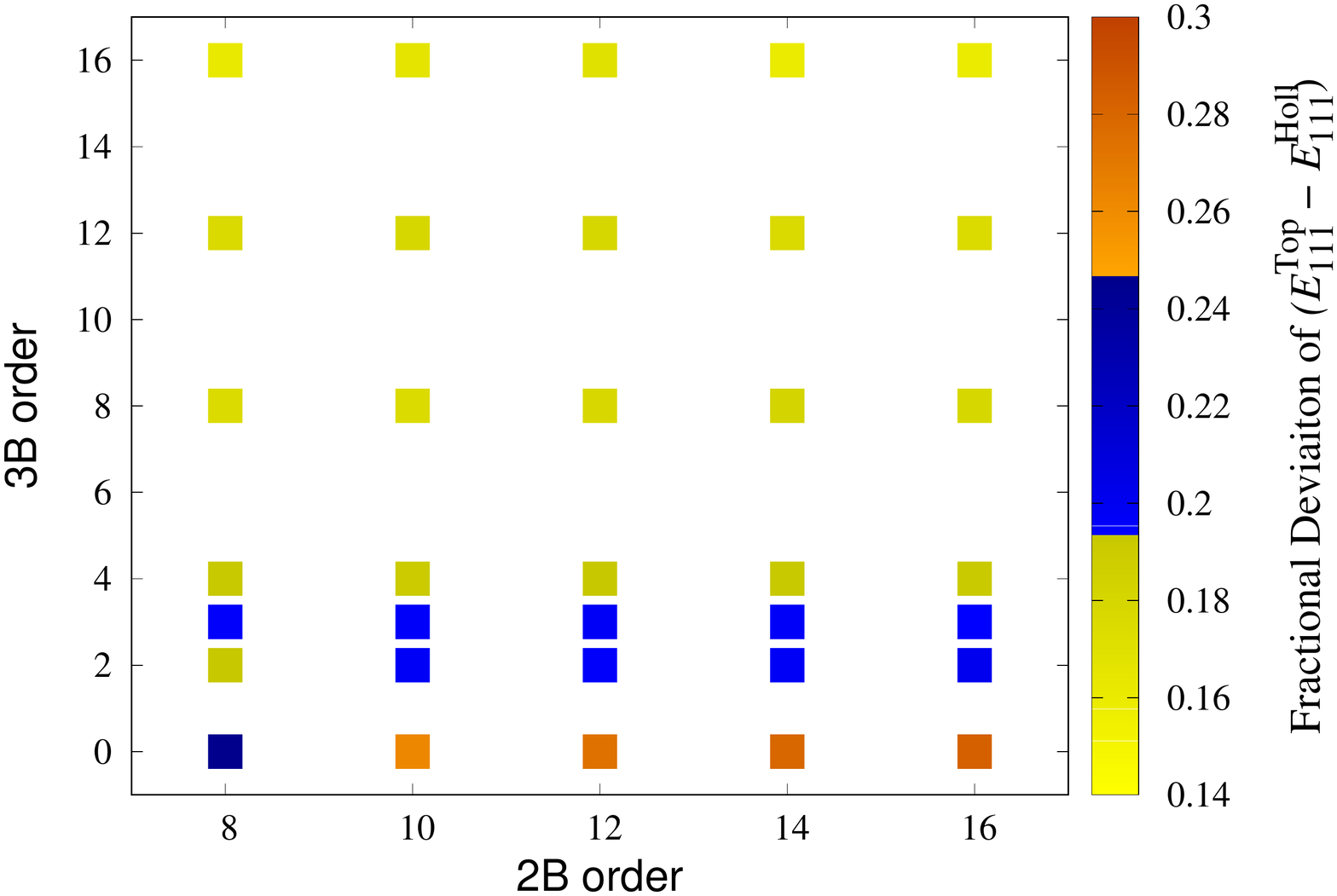}
\caption{\label{fig:order_sweep} Results for sweep of 2-body and 3-body ChIMES polynomial orders, with the same definitions as Figure~\ref{fig:radius_sweep}.}
\end{center}
\end{figure}

Our results indicate weak correlation between choice of 2-body and 3-body polynomial order (Figure~\ref{fig:order_sweep}). Inspection of results from a single 3-body order (e.g., ${{\cal O}_{\rm{3B}}} = 4$) indicates only small changes in any of the validation results presented here with varying values for the 2-body order. We find that 2-body only repulsive energies (with ${{\cal O}_{\rm{3B}}} = 0$) yield some improved accuracy in terms of $\left(E^\mathrm{Top}_{111} -  E^\mathrm{Holl}_{111}\right)$, where we observe values of 0.25 eV for \{${{\cal O}_{\rm{2B}}} = 8$, ${{\cal O}_{\rm{3B}}} = 0$\} and 0.28 eV for \{${{\cal O}_{\rm{2B}}} = 16$, ${{\cal O}_{\rm{3B}}} = 0$\}. However, these 2-body only repulsive energies yield $E_{111}$ surface energies that are consistently over 30\% too high and an $\left(E_{001}/E_{111}\right)$ ratio that is over 20\% too small (e.g., a deviation of 0.020 eV/\AA$^2$ and a ratio of 1.35 for \{${{\cal O}_{\rm{2B}}} = 8$, ${{\cal O}_{\rm{3B}}} = 0$\}). In contrast, we find that increasing the 3-body orders to non-zero values yields substantially improved $E_{111}$ prediction and a somewhat improved $\left(E_{001}/E_{111}\right)$ ratio (a deviation of $<$0.001 eV/\AA$^2$ and ratio of 1.43 for \{${{\cal O}_{\rm{2B}}} = 8$, ${{\cal O}_{\rm{3B}}} = 4$\}). However, large 3-body orders result in somewhat worse values for $\left(E^\mathrm{Top}_{111} -  E^\mathrm{Holl}_{111}\right)$, with values of 0.16 eV for all ${{\cal O}_{\rm{3B}}} = 16$ parameter sets.

Lastly, in order to examine the possibility of including higher-body interactions in $E_\mathrm{Rep}$, we have determined a ChIMES parameterization which includes a 4-body order of four, i.e., \{${{\cal O}_{\rm{2B}}} = 8$, ${{\cal O}_{\rm{3B}}} = 4$, ${{\cal O}_{\rm{4B}}} = 4$\}. We find that this ChIMES model yields virtually identical results to those from \{${{\cal O}_{\rm{2B}}} = 8$, ${{\cal O}_{\rm{3B}}} = 4$\}. For example, both $E^\mathrm{Top}_{111}$ and $E^\mathrm{Holl}_{111}$ agree between both models within 0.01~eV. This is in part due to the fact that the vast majority of the 4-body ChIMES parameters are small and consequently are set to a value of zero by the LASSO/LARS algorithm. This implies that the majority of interactions encompassed in the repulsive energy (for this system, at least) are three-center, only, which can provide guidance for future $E_\mathrm{Rep}$ determinations.

As a result, we opt to proceed with a set of \{${{\cal O}_{\rm{2B}}} = 8$, ${{\cal O}_{\rm{3B}}} = 4$\} for this part of our effort. This set yields a value of $E_{111} = 0.080$, which is nearly identical to the DFT result. In addition, we observe values of $\left(E_{001}/E_{111}\right) = 1.43$ and $\left(E^\mathrm{Top}_{111} -  E^\mathrm{Holl}_{111}\right) = 0.19$~eV. These values are similar to those from \{${{\cal O}_{\rm{2B}}} = 12$, ${{\cal O}_{\rm{3B}}} = 8$\}, but yield slightly improved agreement to DFT results overall (1.70 and 0.68~eV). In addition, the lower polynomial orders help decrease the risk of overfitting that might not be observed with the current validation set.

\subsection{Test of linear least-squares solving methods}
\label{subsec:lsq}

Given our choices of  \{$R^{Ti}_\psi$, $R^{Ti}_n$\} and \{${{\cal O}_{\rm{2B}}}$, ${{\cal O}_{\rm{3B}}}$\}, we now explore several different options for the solving the linear least-squares and regularization to obtain the ChIMES coefficients.  For these tests, we have used the Singular Value Decomposition (SVD) and Least-Angle Regression (LARS) with LASSO regularization (abbreviated as LASSO/LARS). We now offer a brief discussion of each method and leave details to the pertinent references. 

SVD\cite{num_recipes} solves for optimal fitting coefficients directly by performing an eigendecomposition of the generally rectangular $\boldsymbol{A}$ matrix and computing its pseudo-inverse. Regularization can be performed by setting singular values (eigenvalues of the square matrix in the SVD decomposition) with an absolute value below a given threshold to zero. In our work, we take this parameter to be $D_{max} \epsilon$, where $D_{max}$ is the maximum singular value of $\boldsymbol{A}$ and $\epsilon$ is a factor below a value of one. 

LARS is a type of forward step-wise or iterative regression\cite{lars1,lars2}. In this case, all model coefficients are initialized to zero and the covariate (i.e., polynomial values) most correlated to the error residual is determined (i.e., those having the most significant impact on the fit). The corresponding ChIMES parameter is modified incrementally to minimize the error residual until a second covariate yields an equal correlation. At this point, it is included in the active parameter set and both coefficients are modified simultaneously. The process continues until all coefficients are included in the solution, at which point a result equivalent to ordinary least squares fitting is obtained.  In practice, LARS steps using a subset of all possible parameters are chosen.  

LASSO\cite{lasso} is an L1-norm regularization method whereby regularization is based on the sum of the absolute values of the fitting coefficients, which has the effect of shrinking a subset of parameters to zero. In this case, the objective function $F_\mathrm{obj}$ (Equation~\ref{eq:rmse}) is minimized with the following additional constraint:

\begin{equation}
F^{\mathrm{LASSO}}_\mathrm{obj} = F_{\mathrm{obj}} + 2\alpha \sum_{i=1}^{n_i} \left|c_i\right|.
\end{equation}

\noindent Here, $n_i$ is the total number of unique fitting parameters, $c_i$. The parameter $\alpha$ regularizes the magnitude of the fitting coefficients, which reduces possible overfitting.  The LASSO method can be implemented as a variant of LARS where parameters are either added or removed at each solution stage.  We find the LASSO variant of LARS to be numerically stable for ill-conditioned $\boldsymbol{A}$ matrices, which are often found in force matching.

\begin{figure}[!htp]
\begin{center}
\includegraphics[trim=1cm 2cm 1cm 2cm,scale=0.31]{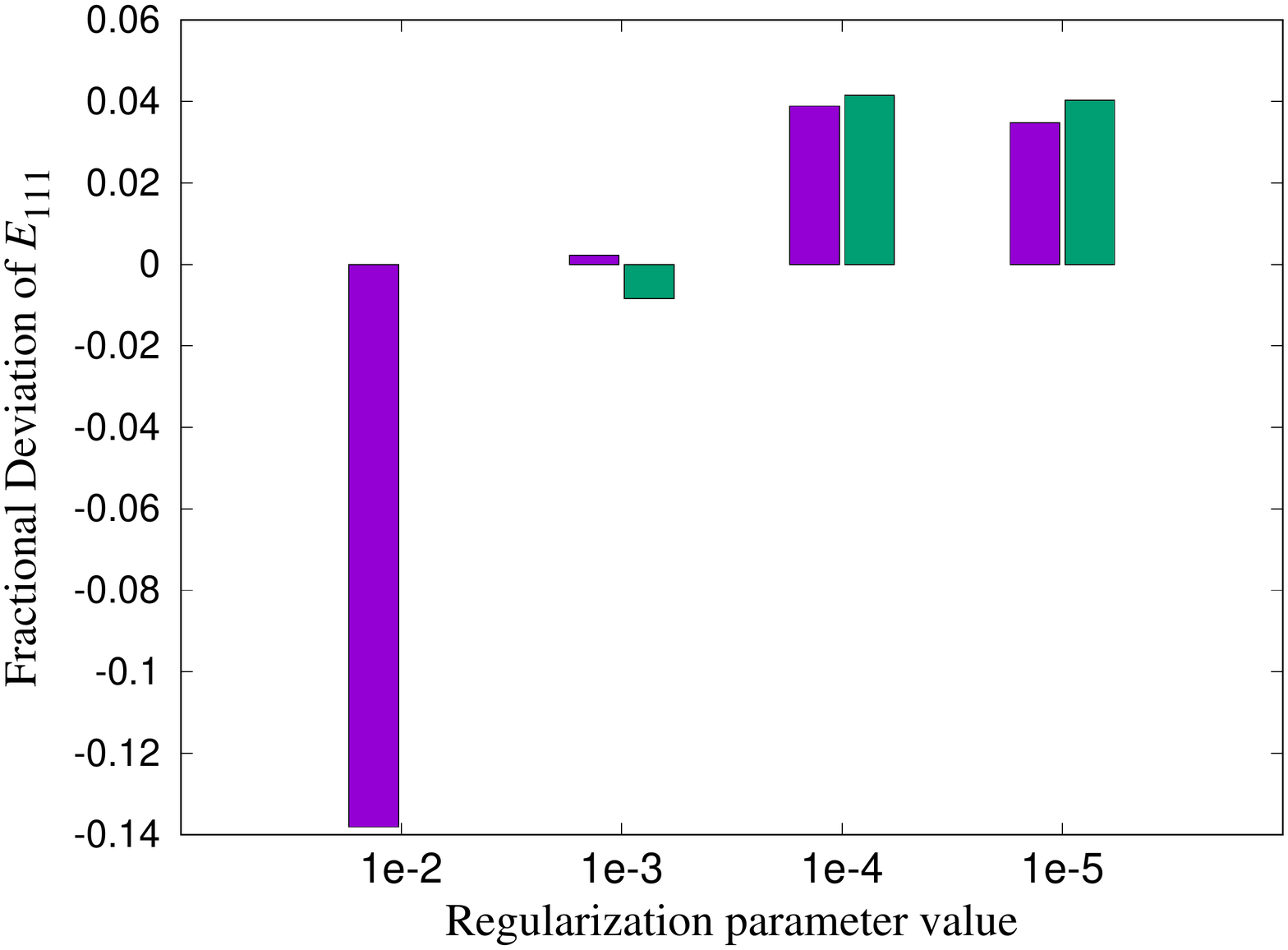}
\includegraphics[trim=1cm 2cm 1cm 2cm,scale=0.31]{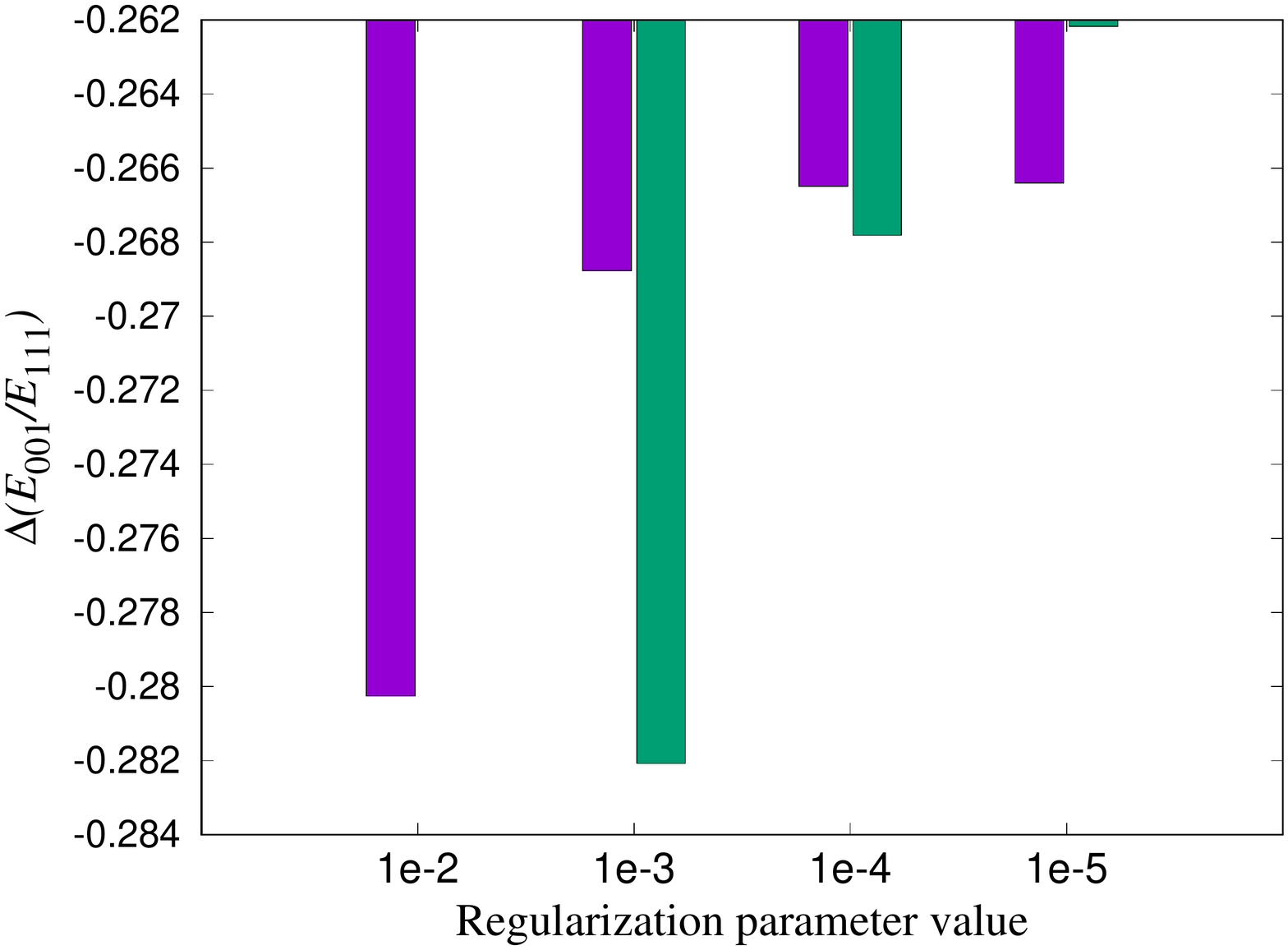}
\includegraphics[trim=1cm 2cm 1cm 2cm,scale=0.31]{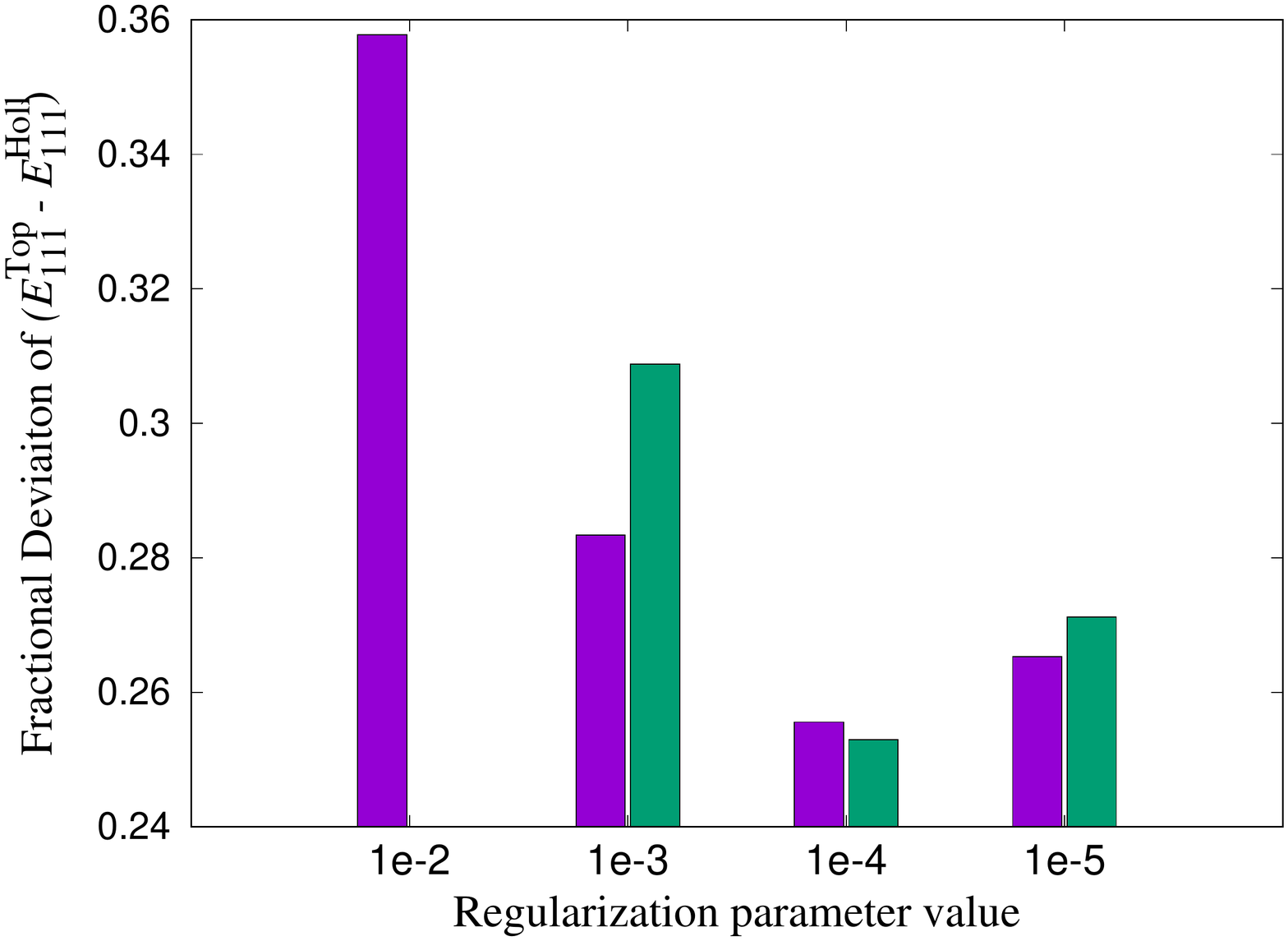}
\caption{\label{fig:lsq_sweep} Results for sweep of different linear least-squares optimization approaches, with the same definitions as Figure~\ref{fig:radius_sweep}. The purple bar corresponds to LASSO/LARS , and the green to SVD. Validation set calculations with an SVD regularization of $10^{-2}$ were unstable and hence are not included.}
\end{center}
\end{figure}

Our exploration spanned SVD $\epsilon$ values between $10^{-5} \leq \epsilon \leq 10^{-3}$ and LASSO/LARS $\alpha$ values between $10^{-5} \leq \alpha \leq 10^{-2}$ (Figure~\ref{fig:lsq_sweep}). \textcolor{black}{Our sweep of LASSO and SVD regularization values and subsequent validation testing help ensure that our DFTB/ChIMES models are not overfit for the conditions of interest here.} SVD optimization with $\epsilon=10^{-2}$ resulted in unstable models and are subsequently not included in our discussion. Results for $E_{111}$ show little dependence on choice of algorithm or regularization parameter. Both optimization algorithms yield very similar values for the $E_{111}$ surface energy, with regularization of $10^{-3}$ showing the best agreement with DFT and LASSO/LARS showing near perfect accuracy. In contrast, the other properties discussed here show some variation between algorithms for regularization of $10^{-3}$. SVD yields a somewhat lower value than LASSO/LARS for the $\left(E_{001}/E_{111}\right)$ ratio (1.418 vs. 1.431) and a slightly higher value for $\left(E^\mathrm{Top}_{111} -  E^\mathrm{Holl}_{111}\right)$ (0.21 vs. 0.19 eV). The agreement overall between algorithms is improved for smaller regularization parameters. LASSO/LARS optimization with $\alpha=10^{-2}$ can appear to be an appealing choice because this parameterization maximizes $\left(E^\mathrm{Top}_{111} -  E^\mathrm{Holl}_{111}\right)$. However, this model yields relatively large errors in the TiH$_2$ lattice constant, (4.311$\,$\AA, compared to a value of 4.440$\,$\AA$\,$ from $\alpha=10^{-3}$ and 4.417$\,$\AA$\,$ from DFT) and $E_{111}$ surface energy (0.092 eV/\AA$^2$, compared to values of 0.080  eV/\AA$^2$ from $\alpha=10^{-3}$ and DFT). Hence, we choose to proceed with LASSO/LARS optimization with $\alpha=10^{-3}$ as the best choice for our TiH$_2$ model, though SVD optimization with $\epsilon=10^{-3}$ is nearly equivalent.

\subsection{Final set of results}

Our final set of hyper-parameter values includes \{$R^{Ti}_\psi=3.6\, \mathrm{au}$, $R^{Ti}_n=5.0\, \mathrm{au}$\} and \{${{\cal O}_{\rm{2B}}} = 8$, ${{\cal O}_{\rm{3B}}} = 4$\}, optimized with LASSO/LARS and regularization of $\alpha=10^{-3}$. This model yields RMS errors of 1.76~eV/\AA$\,$ for hydrogen forces, 1.35~eV/\AA$\,$ for titanium forces, and 0.35~GPa for the stress tensor diagonal. We now discuss results for our full set of validation data. For the remainder of our discussion, we refer to our optimal model as ``DFTB/ChIMES''. 

Results for bulk properties (Table~\ref{tab:bulk_val}) indicate that DFTB/ChIMES yields a lattice constant with errors of only $\sim$0.4\% and 1.0\% from DFT and experiment\cite{tih2_expt_2009}, respectively. However, our model yields a hydrogen bulk vacancy energy ($E_\mathrm{vac}$) that is $\sim$0.5~eV too small. We found that underestimation of the vacancy energy at about this magnitude was typical for all ChIMES parameterizations created in this work. Systematic errors like these could be rectified through a number of methods, such as use of an expanded basis set\cite{Goldman12b}, improved descriptions of charge transfer\cite{Bodrog2012}, or inclusion of multi-center terms in the DFTB Hamiltonian\cite{Goldman15-CPL}, which is the subject of future work.

We find our model yields accurate surface energies for all three low-index facets investigated in this study (Table~\ref{tab:surface_val}). In particular, the $E_{011}$ and $E_{111}$ values are nearly identical to those from DFT. The $E_{001}$ value from DFTB/ChIMES is around 17\% lower than than that for our DFT calculations (0.114 vs. 0.136 eV/\AA$^2$). However, this could be due in part to the internal electric field on the (001) surface configuration studied here. DFTB generally can underestimate surface electrostatic interactions due to its determination of atom-centered point charges only in Coulombic interactions\cite{Witek_DFTB_charges}. 

Calculation of surface and sub-surface hydrogen point vacancies for the (111) and (011) facets shows that DFTB/ChIMES is able to yield the same general trend relative to DFT (Table~\ref{tab:surface_evac}). For the (111) surface, DFTB/ChIMES indicates that removal of a first layer surface hydrogen is 0.52~eV higher than that for the second layer (first sub-surface layer), compared to a value of 0.55~eV from DFT. DFTB/ChIMES essentially recovers the bulk vacancy energy in the second layer, similar to DFT where the second layer vacancy energy is only 0.12~eV greater than that of the bulk. For the (011) surface, DFTB/ChIMES predicts that the first layer $E_\mathrm{vac}$ is 0.18~eV higher than that of the bulk, followed by a 0.57~eV drop in value in the second layer relative to the first. This is in accordance with our DFT results, which indicates an increase of 0.34~eV for the first layer relative to the bulk followed by a relative decrease of 0.51~eV in the second layer. Both DFTB/ChIMES and DFT indicate that the second layer $E_\mathrm{vac}$ is lower than that of the bulk (0.40 eV for DFTB/ChIMES  and 0.17 eV for DFT), and that the bulk $E_\mathrm{vac}$ is recovered by the third surface layer. A value of 0.5~eV could be added as a correction factor to vacancy calculations on these systems.

Our DFTB/ChIMES results show similarly strong agreement with hydrogen surface adsorption energies (Figure~\ref{fig:tih2} and Table~\ref{tab:surface_Hads}). We observe the correct energetic ordering of adsorption on the (111) Top and Hollow sites, though the Hollow site energy is 0.35~eV smaller than that from DFT. We see similar agreement with DFT for the (011) surface. Here, DFTB/ChIMES show close agreement for Top site adsorption with a difference of only 0.05~eV from DFT. Our model yields Bridge-1 and Bridge-2 adsorption energies that differ from DFT by 0.29~eV and 0.21~eV, respectively, and incorrectly predicts that the Top site is the lowest energetically of the three. However, these values are similar in energy for all surface sites and we have overall favorable agreement.

\textcolor{black}{In order to further test the transferability of DFTB/CHIMES, we have performed calculations on both $\alpha$-Ti (hcp lattice) and a gas-phase TiH$_4$ molecule, neither of which were part of our training set (Tables~\ref{tab:a-Ti} and \ref{tab:TiH4}). We find that our DFTB/ChIMES model yields slightly improved accuracy compared to our DFT results for $\alpha$-Ti  with in errors in the lattice constants of only 6.1\% in the $a$ lattice constant and 0.4\% in the $c$ lattice constant relative to experiment\cite{Wyckoff63}. In addition, DFTB/ChIMES predicts the correct tetrahedral structure for TiH$_4$ (H-Ti-H bond angle of 109.47$^\circ$) with a Ti-H bond length of 1.76 \AA, compared to our DFT computed value of 1.71 \AA. Finally, DFTB/ChIMES predicts an energy of reaction for TiH$_4$ $\rightarrow$ TiH$_3$ + H of 2.42 eV, compared to the DFT computed value of 2.88 eV. We note the difference of 0.46 eV nearly exactly corresponds to the systematic error in our hydrogen vacancy calculations.}

\begin{figure}[!htp]
\begin{center}
\includegraphics[scale=0.9]{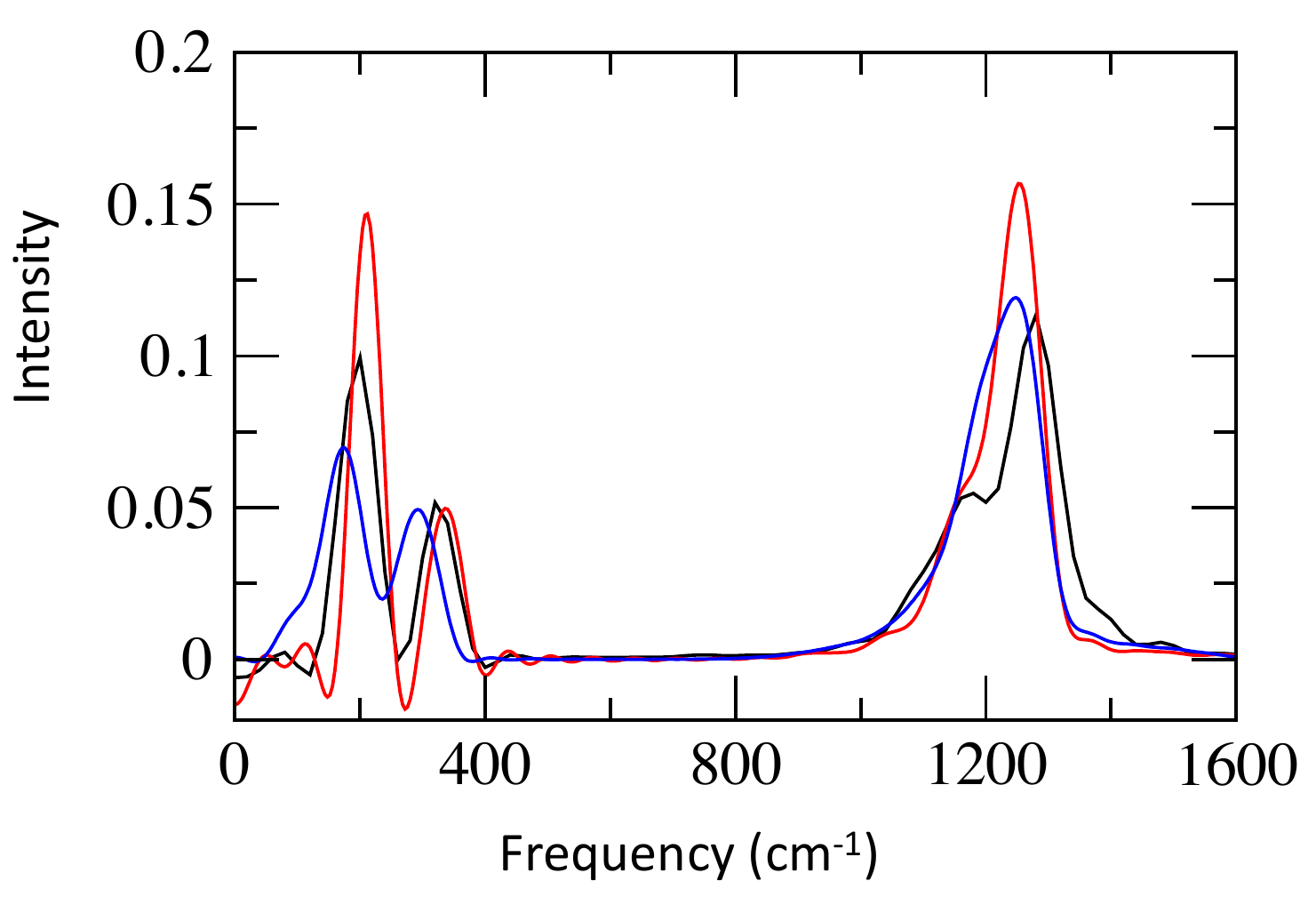}
\caption{\label{fig:vdos} Vibrational density of states calculated from an MD simulation of bulk TiH$_2$ at 400 K. Unit cell simulations using DFT (black) and our DFTB/ChIMES model (red) are compared to results from the periodically replicated supercell (blue). Note that the oscillations at low frequency are artifacts of the Fourier transform for the small timescale and system size of that simulation.}
\end{center}
\end{figure}

Finally, we have computed the vibrational density of states (VDOS) from MD calculations of bulk ground-state TiH$_2$ at 400~K (Figure~\ref{fig:vdos}). We have made comparison to DFTB/ChIMES results using the unit cell and k-point mesh of $10 \times 10 \times 10$, as well as comparison to simulation of a $3 \times 3 \times 2$ supercell (216 atoms total) using a $2 \times 2 \times 2$ k-point mesh. Our results indicate strong agreement between DFTB/ChIMES and DFT for the manifold of both Ti-Ti and H-Ti vibrational modes. In particular, we see a close match between relative peak heights and positions for the DFT and DFTB/ChIMES unit cell simulations at frequencies less than 400 cm$^{-1}$ and greater than 1000 cm$^{-1}$, including the small shoulder at $\sim$1150~cm$^{-1}$. We observe some small shifting of peaks when comparing the DFTB/ChIMES unit cell and replicated supercell results, though these changes are generally $< 10 \mathrm{cm}^{-1}$ and are thus energetically quite small (i.e., $< 10^{-3}$~eV). Overall, we find our DFTB/ChIMES model to yield robust agreement with DFT, allowing for MD calculations with larger system sizes that could incorporate more realistic defect concentrations and/or grain boundaries.
  
\section{Conclusions}

In this work, we have created a workflow for DFTB model creation that allows for nearly exhaustive search for optimal Hamiltonian parameters and repulsive energies. Our efforts leverage the ChIMES reactive force-field, based on linear combinations of Chebyshev polynomials for the determination of the DFTB repulsive energy. Optimal ChIMES parameters can be determined extremely rapidly to create many-body interaction potentials, which facilitates testing of any number of DFTB related parameters. Each design loop is run in a semi-automated fashion, allowing for a systematic exploration of the manifold of options for the confining potential radii, polynomial order and bodied-ness for the ChIMES potential, and type of least-squares algorithm or regularization to solve for the ChIMES parameters. In doing so, we are able to accelerate the down selection of DFTB models for a given application, removing the need for ad hoc parameter choices. 

Our results indicate DFTB/ChIMES models can be accurately determined based on relatively small training data (unit cell MD calculations in this work), even for physically complex systems such as those containing surface chemistry. Further refinement of our TiH$_2$ model could involve inclusion of training data from additional phases and thermodynamic state points. \textcolor{black}{A systematic expansion of our training set could be performed by first generating an approximate model, and then performing DFT calculations on selected snapshots and refitting ChIMES repulsive parameters until convergence is reached\cite{Yoshimoto2006}, similar to our previous force field fitting efforts\cite{Lindsey17}.}  \textcolor{black}{Regardless, our current effort yields accurate results for bulk and surface TiH$_2$ properties, and our model shows strong transferability to bulk $\alpha$-Ti and TiH$_4$.} The small training set could yield significant advantages for computationally challenging systems such as magnetic materials and their interfaces, where DFT data is limited and difficult to generate. Overall, our DFTB/ChIMES approach can have particular impact on myriad of research areas, such as interpretation of imaging and spectroscopy experiments on bulk and interfacial systems, where there is traditionally a strong coupling with atomistic simulation approaches.  

\section*{Acknowledgments}

This work was performed under the auspices of the U.S. Department of Energy by Lawrence Livermore National Laboratory under Contract DE-AC52-07NA27344. Computations were performed at LLNL using the borax, boraxo, quartz, rztrona, and rztopaz massively parallel computers. Projects 18-SI-001 with Jason Jeffries as PI and 20-SI-004 with Brandon Wood as PI were funded by the Laboratory Directed Research and Development Program at LLNL. The ChIMES force field code is available for download at https://github.com/rk-lindsey/chimes\_calculator (last accessed on April 13th, 2021). The DFG grant RTG 2247 is acknowledged. ChIMES potential parameters are available upon request. 

\newpage	

\bibliography{library}

\newpage
\begin{table}[htp]
  \footnotesize
  \caption{Comparison of computed atomic charges for bulk TiH$_2$. Bader charges are computed from DFT whereas Mulliken charges are computed from our DFTB model Hamiltonians.}
  \label{tab:charges}
\begin{tabular}{ccc}
Method & $q_{Ti}$ & $q_{H}$ \\
\hline
\hline
DFT & +1.90 & -0.95 \\
\textcolor{black}{DFTB2-pot} & $-6.9e-3$ & $+3.4e-3$ \\
\textcolor{black}{DFTB2-dens} & +0.32 & -0.16 \\
\textcolor{black}{DFTB3-dens} & +0.430 & -0.215 \\
\end{tabular}
\end{table}

\begin{table}[htp]
  \footnotesize
  \caption{TiH$_2$ bulk properties.}
  \label{tab:bulk_val}
\begin{tabular}{ccc}
Method & Lattice constant (\AA) & H vacancy energy (eV) \\
\hline
\hline
DFTB/ChIMES & 4.400 & 2.47 \\
DFT &  4.417 & 3.00 \\
Expt.\cite{tih2_expt_2009} & 4.45 & -- \\
\\
\end{tabular}
\end{table}

\begin{table}[htp]
  \footnotesize
  \caption{TiH$_2$ surface energies (in eV/\AA$^2$).}
  \label{tab:surface_val}
\begin{tabular}{ccc}
Surface & DFTB/ChIMES & DFT \\
\hline
\hline
111& 0.080 & 0.080 \\
011& 0.105 & 0.101 \\
001& 0.114 &  0.136 \\
\end{tabular}
\end{table}

\begin{table}[!htp]
  \footnotesize
  \caption{TiH$_2$ surface vacancy energies (in eV). The ``layer'' label corresponds to the surface layer from which the H atom is removed. }
  \label{tab:surface_evac}
\begin{tabular}{cccc}
Surface & Layer & DFTB/ChIMES & DFT \\
\hline
\hline
111& 1st & 2.98 & 3.67 \\
& 2nd & 2.46 & 3.12 \\
& 3rd & 2.46 & 2.97 \\
& 4th & 2.49 & 3.02 \\
011& 1st & 2.65 & 3.34 \\
& 2nd & 2.07 & 2.83 \\
& 3rd & 2.54 & 2.99 \\
& 4th & 2.52 & 3.00 \\

\end{tabular}
\end{table}

\begin{table}[!htp]
  \footnotesize
  \caption{Surface hydrogen adsorption energies on TiH$_2$ surface sites (in eV).}
  \label{tab:surface_Hads}
\begin{tabular}{cccc}
Surface & Site & DFTB/ChIMES & DFT \\
\hline
\hline
111& Top & -1.888 & -1.760 \\
& Hollow & -2.081 & -2.440  \\
011& Top & -2.383 & -2.332  \\
& Bridge-1 & -2.154 & -2.442 \\
& Bridge-2 & -2.132 & -2.342 \\
\end{tabular}
\end{table}

\begin{table}[htp]
  \footnotesize
 \caption{\textcolor{black}{$\alpha$-Ti lattice constants in \AA.}}
  \label{tab:a-Ti}
 {\color{black}\begin{tabular}{ccc}
Method & $a$ & $c$ \\
\hline
\hline
DFTB/ChIMES & 3.13 & 4.67 \\
DFT &  2.82 & 4.55 \\
Expt.\cite{Wyckoff63} & 2.95 & 4.69 \\
\\
\end{tabular}}
\end{table}

\begin{table}[htp]
  \footnotesize
 \caption{\textcolor{black}{TiH$_4$ gas-phase molecular data.}}
  \label{tab:TiH4}
 {\color{black}\begin{tabular}{cccc}
Method & H-Ti-H bond angle (degrees) & Ti-H bond length (\AA) & H dissociation energy (eV) \\
\hline
\hline
DFTB/ChIMES & 109.47 & 1.76 & 2.42 \\
DFT &  109.47 & 1.71 & 2.88 \\
\\
\end{tabular}}
\end{table}

\newpage

\newpage

\newpage

\newpage

\newpage

\newpage

\newpage

\begin{figure}
\begin{center}
\includegraphics[scale=0.5]{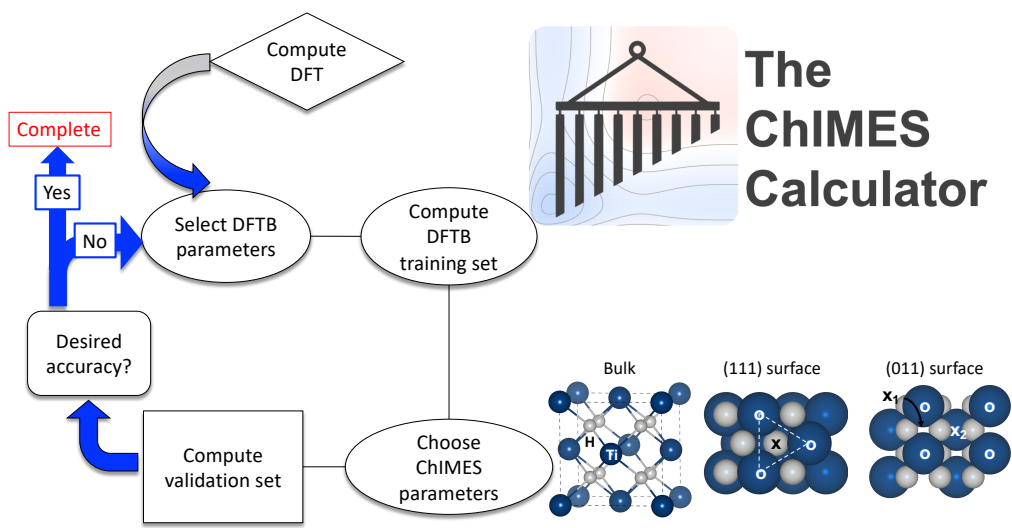}
\caption*{\textbf{TOC Graphic}}
\end{center}
\end{figure}

\end{document}